\documentclass[a4paper,11pt]{article}
\pdfoutput=1 

\usepackage{jcappub} 

\usepackage[T1]{fontenc} 
\usepackage{amsmath}
\usepackage{comment}
\usepackage{cleveref}
\usepackage{subcaption}

\title{\boldmath Induced gravitational waves: the effect of first order tensor perturbations}

\newcommand{\refrep}[1]{\textcolor{black}{#1}}
\newcommand{\refrepencore}[1]{\textcolor{black}{#1}}
\newcommand{\cirpol}[1]{\textcolor{black}{#1}}

\author[a]{Rapha{\" e}l Picard}
\author[a]{and Karim A.~Malik}

\affiliation[a]{Department of Physics and Astronomy, Queen Mary University of London, UK}

\emailAdd{r.h.j.picard@qmul.ac.uk}
\emailAdd{k.malik@qmul.ac.uk}

\abstract{Scalar induced gravitational waves contribute to the cosmological gravitational wave background. They can be related to the primordial density power spectrum produced towards the end of inflation and therefore are a convenient new tool to constrain models of inflation. These waves are sourced by terms quadratic in perturbations and hence appear at second order in cosmological perturbation theory. While the focus of research so far was on purely scalar source terms we also study the effect of including first order tensor perturbations as an additional source. This gives rise to two additional source terms: a term  quadratic in the  tensor perturbations and a cross term involving mixed scalar and tensor perturbations. We present full analytical expressions for the spectral density of these new source terms and discuss their general behaviour. To illustrate the generation mechanism we study two toy models containing a peak on small scales. For these models we show that the scalar-tensor contribution becomes non-negligible compared to the scalar-scalar contribution on smaller scales. We also consider implications for future gravitational wave surveys.}

\begin{document}
\maketitle
\flushbottom
\newpage
\section{Introduction}
\label{sec:intro}
The first direct detection of gravitational waves by the LIGO/Virgo collaboration \cite{LIGOScientific:2016aoc}, resulting from the merger of two black holes, and subsequent detections, reignited cosmologists interests in using gravitational waves as a novel probe into the early universe. Recently, thanks to \refrep{pulsar} timing arrays, the NANOGrav collaboration \cite{NANOGrav:2023gor}, the EPTA/InPTA \cite{Antoniadis:2023utw, Antoniadis:2023rey, Antoniadis:2023zhi}, the CPTA \cite{Xu:2023wog} and the PPTA \cite{Reardon:2023gzh, Zic:2023gta} published data that provides evidence for a low frequency stochastic gravitational wave background (SGWB). Although the source of the SGWB is as yet undetermined, this evidently is a milestone for researchers who have predicted a SGWB of cosmological origin. The NANOGrav collaboration also fitted cosmological sources \cite{NANOGrav:2023hvm} such as domain walls, cosmic strings and cosmological phase transitions to their data release, but stress that their results should not (yet) be taken as evidence for new physics. 

On the theory side, inflation allows for two different and independent ways of production of gravitational waves, leading to different observational signatures. At linear order, gravitational waves produced during inflation have been studied for some time now and have become a well established tool to discriminate between models of inflation \cite{Liddle:2000cg}. They can leave a distinct imprint in the Cosmic Microwave Background (CMB) in form of B-mode polarization. For example, the Planck collaboration has extracted an upper limit on the tensor-to-scalar ratio $r<0.065$, at $k=0.002\text{ Mpc}^{-1}$\cite{Planck:2018vyg}. Going beyond linear order in perturbation theory allows us to study scalar induced gravitational waves (SIGWs), which are gravitational waves that are sourced by terms quadratic in linear order scalar quantities. SIGWs were first studied in the $1960's$ by K.~Tomita \cite{Tomita:1967} and later on by Refs.~\cite{Matarrese:1992rp, Matarrese:1993zf, Matarrese:1997ay, Ananda:2006af, Baumann:2007zm}, amongst others. Unlike first order gravitational waves, they have a non-negligible source term which is related to the primordial scalar power spectrum. These fluctuations couple to produce gravitational waves. The  Planck collaboration determined the scalar power spectrum and gives the value of the spectral index $n_s=0.9626\pm 0.0057$ at $k=0.05\text{ Mpc}^{-1}$\cite{Planck:2018vyg}. 


In order for SIGWs to be detectable there needs to be some enhancement of the scalar power spectrum (see e.g.~Ref.~\cite{Domenech:2021ztg} for a review). Both the parameters $r$ and $n_s$ are constrained at CMB scales, and \textit{a priori} there is no reason  that these values hold for scales that exit the horizon in the last stages of inflation. There are multiple ways to enhance the primordial scalar power spectrum: for example in single field inflation via a period of ultra-slow-roll \cite{Kinney:1997ne, Iacconi:2021ltm, Motohashi:2017kbs, Garcia-Bellido:2017mdw} or in multi-field models of inflation \cite{Fumagalli:2020adf, Palma:2020ejf, Braglia:2020eai}. The basic idea is that the power spectrum of SIGWs ($\mathcal{P}_{SS}$) is related to the linear power spectrum, $\mathcal{P}_{\zeta}$, as: $\mathcal{P}_{ss} \propto \mathcal{P}_{\zeta}^2$, so that if there is an enhancement/peak in $\mathcal{P}_{\zeta}$ there will also be one in $\mathcal{P}_{ss}$. Depending on the scale at which this enhancement happens, the resulting background of gravitational waves can be in principle observable by future detectors such as the Laser Interferometer Space Antenna (LISA) \cite{Bartolo:2016ami},  DECi-hertz Interferometer Gravitational wave Observatory (DECIGO) \cite{Kawamura:2020pcg} and the Einstein Telescope (ET) \cite{Hild:2010id} to name a few.

Most recent research has been centered around how scalar fluctuations couple to produce SIGWs. However, there is also the possibility that a scalar fluctuation couples with a tensor mode or that two tensor modes couple to induce more general second order gravitational waves (SOGWs). Moreover, if there is also a peak (on smaller scales) of primordial gravitational waves, then we could expect to see its imprint also at second order. Multifield models of inflation can produce such a peak, for example models containing a spectator axion coupled to a $SU(2)$ gauge field \cite{Barnaby:2011qe, Thorne:2017jft, Dimastrogiovanni:2016fuu}. This was looked at in Ref.~\cite{Chang:2022vlv} for monochromatic power spectra and in Ref.~\cite{Gong:2019mui} analytical expressions for the kernels are derived. Recently, Ref.~\cite{Bari:2023rcw} studied scalar-tensor induced gravitational waves in a radiation dominated universe with a possibility of having an initial parity violation and Ref.~\cite{Yu:2023lmo} looked at the case where the primordial power spectrum follows a sudden-broken power-law.

In this paper we look at the production of SOGWs in a homogeneous and isotropic universe sourced by first order scalar and tensor modes. We will derive the power spectrum of SOGWs $\mathcal{P}_{h^{(2)}}(\eta, k)$ defined as
\begin{equation}\label{2pointfunction}
    \langle h^{(2)}_{\lambda}(\eta, \mathbf{k})h^{(2)}_{\lambda ^{\prime}}(\eta, \mathbf{k^{\prime}}) \rangle =  \delta ^{(3)} (\mathbf{k} + \mathbf{k^{\prime}}) \delta ^{\lambda \lambda ^{\prime}} \frac{2\pi ^2}{k^3} \mathcal{P}_{h^{(2)}}(\eta, k)\,,
\end{equation}
where $\lambda,\lambda ^{\prime} = R,L$ are the polarization states of the GWs. We will use the fact that the spectrum can be decomposed into  the sum of the independent contributions of the scalar-scalar, tensor-tensor ($\mathcal{P}_{tt}$) and scalar-tensor induced waves ($\mathcal{P}_{st}$) power spectra, such that: 
\begin{equation}
    \mathcal{P}_{h^{(2)}}(\eta, k) = \mathcal{P}_{ss}(\eta, k)+\mathcal{P}_{tt}(\eta, k)+\mathcal{P}_{st}(\eta, k) \,.
\end{equation}
Assuming there is a peaked power spectrum for both first the order power spectrum of scalars and tensors, we will parameterise them by four parameters: $\mathcal{A}_{\zeta}$, the amplitude of the scalar power spectrum, $k_{\zeta}$, the scale at which the scalar power spectrum peaks), $\mathcal{A}_{h}$, the amplitude of the tensor power spectrum and $k_h$ the scale at which the tensor power spectrum peaks.
Given these parameters we will show that there is a non-negligible contribution from considering tensors as an additional source term. Our results include the \refrep{spectral} density of SOGWs and expressions for their corresponding kernels. 

We begin in Section \ref{section2} by investigating what terms source gravitational waves and solve the equation of motion for these induced tensor modes. Section \ref{section3} shows how to obtain the power spectrum of SOGWs, followed by Section \ref{section4} where we derive expressions for the spectral density and plot the latter for peaked power spectras. We use the `mostly plus' metric, Greek indices run over from $0,1,2,3$, whilst Latin indices run over the spatial coordinates $1,2,3$ and we work in natural units, i.e.~$c=1$, and define $M_{Pl}^2=(8\pi G)^{-1}$.
\section{The second order gravitational wave equation}\label{section2}
\subsection{Extracting gravitational wave source terms}
We use the conformal Newtonian gauge in a flat FLRW background, see e.g.~Ref.~\cite{Malik:2008im} for a review of cosmological perturbation theory. We do not include vectors at any order since they are diluted away during inflation. The line element is given by \cite{Peter:2013avv}
\begin{equation}
    ds^2 = a^2(\eta)\left [-(1+2\Phi + \Phi ^{(2)} )d\eta ^2+\left ((1-2\Psi - \Psi ^{(2)})\delta _{ij}  +2\overline{h}_{ij}+\overline{h}^{(2)}_{ij} \right )dx^idx^j \right]\,,
\end{equation}
where $\eta$ is conformal time, $a(\eta)$ the conformal scale factor and $\delta _{ij}$ is the Kronecker delta. Metric perturbations $\Phi ^{(n)} , \Psi ^{(n)}$ are respectively the lapse and curvature perturbations and $\overline{h}_{ij}^{(n)}$ are tensorial perturbations, which are transverse-traceless (TT): $\partial ^i \overline{h}^{(n)}_{ij}=\delta ^{ij}\overline{h}^{(n)}_{ij}=0$. Here $(n)$ is the perturbative order. Additionally, to reduce clutter, we have taken the convention to only show the order of perturbation for second order variables and get rid of the overline which indicates when a tensorial quantity is TT.

We are interested in the second order equation of motion for tensors, therefore only TT tensorial quantities will contribute. We can extract TT quantities by applying the TT projection operator $\Lambda ^{ij}_{ab}$ to the Einstein equations at second order. This operator can be defined in the following way, see Ref.~\cite{Domenech:2020ssp},
\begin{equation}\label{TTprojector}
    \Lambda ^{ij}_{ab} = \left (\delta ^i_a - \frac{\partial ^i \partial _a}{\nabla ^2} \right )\left (\delta ^j_b - \frac{\partial ^j \partial _b}{\nabla ^2} \right ) -\frac{1}{2}\left (\delta _{ab} - \frac{\partial _a \partial _b}{\nabla ^2}\right )\left (\delta ^{ij} - \frac{\partial ^i \partial ^j}{\nabla ^2} \right )\,,
\end{equation}
where the inverse Laplacian defined by its action on the Laplacian of a tensor field $X$ as: $\nabla ^{-2}(\nabla ^2 X) = X$.  The second order Einstein equations can be found in App.~\ref{2oEinstein}. Hence, in real space we find
\begin{equation}\label{realspaceSOGWs}
    \Lambda ^{ij}_{ab} G_{ij}^{(2)} = \frac{1}{M_{Pl}^2}\Lambda ^{ij}_{ab} T_{ij}^{(2)}\,. 
\end{equation}
We study SOGWs sourced by modes produced during inflation and re-entering during the radiation dominated (RD) epoch, and describe the energy content of the universe as an adiabatic perfect fluid with energy momentum-tensor
\begin{equation}\label{EMtensor}
    T_{\mu \nu} = (\rho + P ) u_{\mu} u_{\nu} + P g_{\mu \nu}\,,
\end{equation}
where $\rho$ and $P$ are, respectively, the energy density and pressure, and $u^{\mu}$ the 4-velocity. Throughout this paper we neglect anisotropic stress, as it has been shown to have a negligible effect on the spectrum of SIGWs \cite{Baumann:2007zm}. Then Eq.~(\ref{realspaceSOGWs}) becomes:
\begin{equation} 
\label{GW2eq}
    h_{ab}^{\prime \prime (2)} + 2\mathcal{H}h_{ab}^{\prime (2)} - \nabla ^2h_{ab}^{(2)}= \Lambda ^{ij}_{ab}S_{ij}\,,
\end{equation}
where $S_{ab}$ is the source term. We separate the source term into quadratic pure first order scalar contributions (ss), quadratic pure first order tensor contributions (tt) and a scalar-tensor cross term (st), such that:
\begin{equation}
    S_{ij}=S_{ij}^{ss} + S_{ij}^{tt} + S_{ij}^{st} \,.
\end{equation} 
These terms are given by:
\begin{subequations}\label{sourceterms}
\begin{align}
\label{scalarsource}
S^{ss} _{ij} & = \frac{8}{3(1+w)}\left [(\partial _i\Psi +\frac{\partial _i \Psi ^{\prime}}{\mathcal{H}})(\partial _j \Psi +\frac{\partial _j \Psi ^{\prime}}{\mathcal{H}})\right ] +4 \partial _i \Psi \partial _j \Psi \,,
\\
\label{tensorsource}
\begin{split}
    S^{tt}_{ij} & = - 4h^{cd} \partial _c \partial _d h_{ij} + 4\partial _d h_{jc} \partial ^c h_i^d
    - 4\partial _d h_{jc} \partial ^d h^c_i + 8 h^{dc} \partial _i \partial _ c h_{jd} \\& 
    +4h^{c\prime}_i h_{jc}^{\prime}+ 2 \partial _i h^{cd} \partial _j h_{cd}\,,
\end{split}
\\
\label{sclartensorsource}
\begin{split}
    S^{st}_{ij} & = 8\Psi\nabla^2h_{ij} + 8\partial _ch_{ij}\partial ^c \Psi + 4h_{ij} (   \mathcal{H}(1+3c_s^2) \Psi^{\prime} +(1-c_s^2)\nabla^2\Psi )\,.
\end{split}
\end{align}
\end{subequations}
Note that these expressions have been simplified using first order equations of motion, which can be found 
in App.~\ref{1oEinstein}.
Now that we have extracted the source terms we can move on to solving Eq.~(\ref{GW2eq}), which we will do in the next section.
\subsection{Solving the GW equation}
It is convenient to solve this equation in Fourier space, see App.~\ref{AppendixPol} for conventions. For each polarization \cirpol{$\lambda=R,L$}, Eq.~(\ref{GW2eq}) turns into
\begin{equation} \label{GW2fourier}
    h_{\lambda}^{(2)\prime \prime} (\eta , \mathbf{k}) + 2\mathcal{H}h_{\lambda}^{(2)\prime} (\eta , \mathbf{k}) +k^2h_{\lambda}^{(2)} (\eta , \mathbf{k}) =  4\mathcal{S}_{\lambda} (\eta , \mathbf{k})\,,
\end{equation}
with $S_{\lambda}(\eta , \mathbf{k})=S_{\lambda}^{ss}(\eta , \mathbf{k}) + S_{\lambda}^{tt}(\eta , \mathbf{k}) + S_{\lambda}^{st}(\eta , \mathbf{k})$. To solve the above, we use Green's method:
\begin{equation} \label{GW2sol}
    h^{(2)}_{\lambda}(\eta , \mathbf{k}) = \frac{4}{a(\eta)} \int ^{\eta} d\overline{\eta} G_{\mathbf{k}} (\eta ,\overline{\eta} )a(\overline{\eta}) S_{\lambda}(\overline{\eta},\mathbf{k})\,,
\end{equation}
where $G_{\mathbf{k}} (\eta ,\overline{\eta} )$ solves:
\begin{equation}
   G_{\mathbf{k}}^{\prime \prime}  (\eta ,\overline{\eta} ) + 2\mathcal{H} G_{\mathbf{k}}^{\prime }(\eta ,\overline{\eta} )+k^2  G_{\mathbf{k}}(\eta ,\overline{\eta} ) = \delta (\eta - \overline{\eta})\,.
\end{equation}
In our work we look at modes which re-enter the horizon and interact during RD, hence $\mathcal{H}=1/\eta$, which leads to: 
\begin{equation}
    G_{\mathbf{k}}(\eta , \overline{\eta}) = \frac{1}{k}\sin{(k\eta -k\overline{\eta})}\,.
\end{equation}
The solution of Eq.~(\ref{GW2sol}) depends on the evolution of the first order scalar and tensor modes. When these modes are super-Hubble their amplitudes are constant (i.e. when $k\eta<1$) and set during inflation. This implies we can split their respective equations of motion as a combination of its primordial value and a transfer function. For the first order scalar modes we define
\begin{equation}\label{scalartransfer}
    \Psi (\eta , \mathbf{k}) = \Psi _{\mathbf{k}}\frac{3}{(c_sx)^2} \left (\frac{\sin{c_sx}}{c_sx} - \cos{c_sx} \right ) \equiv \frac{2}{3} \mathbf{\zeta}_{\mathbf{q}} T_{\Psi}(c_sx)\,,
\end{equation}
where $\zeta$ is the comoving curvature perturbation conserved when the mode is super-Hubble. For the first order tensor modes: 
\begin{equation}\label{tensortransfer}
    h^{\lambda} (\eta , \mathbf{q}) = h^{\lambda}_{\mathbf{k}} \frac{\sin{x}}{x} \equiv h^{\lambda}_{\mathbf{k}} T_h(x) \,,
\end{equation}
where we have defined $x=k\eta$ in order to solve Eq.~(\ref{firstorderscalareom}) and Eq.~(\ref{firstordertensoreom}) in Fourier space. Both transfer functions are \cirpol{even under  $k\rightarrow -k$}.

Finally, splitting the contributions to Eq.~(\ref{GW2sol}) into pure scalar, tensor and mixed we arrive at using Eqs.~(\ref{sourceterms}), (\ref{scalartransfer}) and (\ref{tensortransfer}):
\begin{subequations}\label{GWexpression}
\begin{align}
h_{\lambda ,ss}^{(2)} &= \frac{16}{9}\int \frac{d^3\mathbf{p}}{(2\pi)^{\frac{3}{2}}} q^{\lambda}_{ss}(\mathbf{k},\mathbf{p}) \zeta_{\mathbf{p}} \zeta_{\mathbf{k}-\mathbf{p}}  \int ^{\eta} _0 d\overline{\eta} \frac{a(\overline{\eta})}{a(\eta)}G_{\mathbf{k}} (\eta ,\overline{\eta} )f^{ss}(c_s\overline{\eta} p,c_s\overline{\eta} |\mathbf{k}-\mathbf{p}|)\,, 
\\
\begin{split}
h_{\lambda ,tt}^{(2)} &=\cirpol{\sum _{\lambda _1 , \lambda _2} } 4\int \frac{d^3\mathbf{p}}{(2\pi)^{\frac{3}{2}}} q^{\lambda , \lambda _1 , \lambda _2}_{tt,1}(\mathbf{k},\mathbf{p}) h_{\mathbf{p}}^{\lambda _1} h_{\mathbf{k}-\mathbf{p}}^{\lambda _2}  \int ^{\eta} _0 d\overline{\eta}\frac{a(\overline{\eta})}{a(\eta)}G_{\mathbf{k}} (\eta ,\overline{\eta} ) f_1^{tt}(\overline{\eta} p,\overline{\eta} |\mathbf{k}-\mathbf{p}|) \\
&+\cirpol{\sum _{\lambda _1 , \lambda _2} }4\int \frac{d^3\mathbf{p}}{(2\pi)^{\frac{3}{2}}} q^{\lambda , \lambda _1 , \lambda _2}_{tt,2}(\mathbf{k},\mathbf{p}) h_{\mathbf{p}}^{\lambda _1} h_{\mathbf{k}-\mathbf{p}}^{\lambda _2}  \int ^{\eta} _0 d\overline{\eta}\frac{a(\overline{\eta})}{a(\eta)}G_{\mathbf{k}} (\eta ,\overline{\eta} ) f_2^{tt}(\overline{\eta} p,\overline{\eta} |\mathbf{k}-\mathbf{p}|)\,,
\end{split}
\\
\begin{split}
h_{\lambda ,st}^{(2)} &=\cirpol{\sum _{\lambda _3} } \frac{8}{3} \int \frac{d^3\mathbf{p}}{(2\pi)^{\frac{3}{2}}} q^{\lambda , \lambda _3}_{st}(\mathbf{k},\mathbf{p}) h_{\mathbf{p}}^{\lambda _3} \zeta_{\mathbf{k}-\mathbf{p}} \int ^{\eta} _0 d\overline{\eta}\frac{a(\overline{\eta})}{a(\eta)}G_{\mathbf{k}} (\eta ,\overline{\eta} ) f^{st}(\overline{\eta} p,c_s\overline{\eta} |\mathbf{k}-\mathbf{p}|)\,.
\end{split}
\end{align}
\end{subequations}

We note that there is a sum over \cirpol{$\lambda_{1},\lambda_{2},\lambda_{3}=R,L$} (i.e. for each first order gravitational wave source terms there is a sum over the two polarizations). Each contribution (scalar-scalar, tensor-tensor and scalar-tensor) has been split into three separate parts: First, the functions $q^{\lambda}(\mathbf{k},\mathbf{p})$ encoding the polarization, shown below:
\begin{subequations}\label{defpoldecomp}
\begin{align}
q^{\lambda }_{ss}(\mathbf{k},\mathbf{p}) &= \cirpol{\left ( q^{ab}_{\lambda}(\mathbf{k}) \right ) ^*}  p_{a}p_{b}\,, \\
\label{polarazationdeftt1}
\begin{split}
q^{\lambda , \lambda _1 , \lambda _2}_{tt,1}(\mathbf{k},\mathbf{p}) &= \cirpol{\left ( q^{ab}_{\lambda}(\mathbf{k}) \right ) ^*} \bigg(   q_{ab}^{\lambda _2}(\mathbf{p}) q^{cd}_{\lambda _1} (\mathbf{k}-\mathbf{p})p_cp_d -   q_{bc}^{\lambda _1}(\mathbf{k}-\mathbf{p}) q^{d}_{a,\lambda _2} (\mathbf{p})(k-p)_dp^c  \\
&+ q_{bc}^{\lambda _1}(\mathbf{k}-\mathbf{p}) q^{c}_{a,\lambda _2} (\mathbf{p})(k-p)_dp^d -2 q^{dc}_{\lambda _1}(\mathbf{k}-\mathbf{p}) q^{\lambda _2}_{bd} (\mathbf{p})p_ap_c \\
&-\frac{1}{2} q^{cd}_{\lambda _1}(\mathbf{k}-\mathbf{p}) q^{\lambda _2}_{cd}(\mathbf{p})(k-p)_ap_b \bigg )\,,
\end{split}
\\
\label{polarazationdeftt2}
q^{\lambda , \lambda _1 , \lambda _2}_{tt,2}(\mathbf{k},\mathbf{p}) &= \cirpol{\left ( q^{ab}_{\lambda}(\mathbf{k}) \right ) ^*} q_{a}^{c,\lambda _1}(\mathbf{k}-\mathbf{p}) q^{\lambda _2}_{bc} (\mathbf{p})\,. \\
q^{\lambda , \lambda _3}_{st}(\mathbf{k},\mathbf{p}) &= \cirpol{\left ( q^{ab}_{\lambda}(\mathbf{k}) \right ) ^*}q^{\lambda _3}_{ab}(\mathbf{p})\,,
\end{align}
\end{subequations}
\refrep{where $\cirpol{\left ( q^{ab}_{\lambda}(\mathbf{k}) \right ) ^*}$ is the polarization tensor of the induced wave and comes from the transverse-traceless projector in Fourier space, see Refs.~\cite{Ananda:2006af, Baumann:2007zm}}. The second part is the primordial values of the perturbations $\mathbf{\zeta}$ and $h^{\lambda}$. The third part of Eq.~(\ref{GWexpression}) is an integral over the retarded time $\overline{\eta}$ of the Green's and transfer functions which describes how modes behave once they re-enter the horizon:
\begin{subequations}\label{sourcetermsfourier}
\begin{align}
\begin{split}
    f^{ss}(c_s\eta p,c_s\eta|\mathbf{k}-\mathbf{p}|) &= \frac{2}{3(1+w)} \bigg [ \bigg ((T_{\Psi}(c_s\eta |\mathbf{k}-\mathbf{p}|) + \frac{T^{\prime}_{\Psi}(c_s\eta |\mathbf{k}-\mathbf{p}|)}{\mathcal{H}}\bigg )  \bigg ( T_{\Psi}(c_s\eta p) \\
    &+ \frac{T^{\prime}_{\Psi}(c_s\eta p)}{\mathcal{H}} \bigg )  \bigg ] 
    +T_{\Psi}(c_s\eta |\mathbf{k}-\mathbf{p}|)T_{\Psi}(c_s\eta p)\,, 
\end{split}
\\
f^{tt}_1(\eta p,\eta|\mathbf{k}-\mathbf{p}|) &=T_h(\eta|\mathbf{k}-\mathbf{p}|)T_h(\eta p)\,,\\
f^{tt}_2(\eta p,\eta|\mathbf{k}-\mathbf{p}|) &= T^{\prime}_h(\eta|\mathbf{k}-\mathbf{p}|)T^{\prime}_h(\eta p)\,,\\
\begin{split}
     f^{st}(\eta p,c_s \eta|\mathbf{k}-\mathbf{p}|) &= -2p^2T_h(\eta p) T_{\Psi}(c_s \eta|\mathbf{k}-\mathbf{p}|)\refrep{-2(\mathbf{k}-\mathbf{p})\cdot \mathbf{p}}T_h(\eta p) T_{\Psi}(c_s \eta|\mathbf{k}-\mathbf{p}|) \\
    &+ \mathcal{H}(1+3c_s^2)T_h(\eta p) T_{\Psi}^{\prime}(c_s \eta|\mathbf{k}-\mathbf{p}|)-|\mathbf{k}-\mathbf{p}|^2(1-c_s^2)T_h(\eta p) \\
    &\times T_{\Psi}(c_s \eta|\mathbf{k}-\mathbf{p}|)\,. 
\end{split}
\end{align}
\end{subequations}
Note that the source terms in Fourier space are given by a convolution of the first order perturbations, hence there is a freedom in relabelling (i.e. a symmetry between) $\mathbf{p}\leftrightarrow \mathbf{k}-\mathbf{p}$. This symmetry will be used to simplify calculations when considering the two-point correlation function of $ h_{\lambda}^{(2)}$, in the following section.
\section{Extracting the power spectrum}\label{section3}
To extract the power spectrum we substitute Eq.~(\ref{GW2sol}) into the left-hand side of Eq.~(\ref{2pointfunction}). By factorizing out each terms primordial value in Eq.~(\ref{GWexpression}), it becomes more apparent what contributes to the final power spectrum:
\begin{subequations}
    \begin{align}
        h^{\lambda (2)}_{ss}(\eta, \mathbf{k}) &= \frac{16}{9\refrep{k^3}\eta} \int \frac{d^3\mathbf{p}}{(2\pi)^{\frac{3}{2}}} q^{\lambda}_{ss} (\mathbf{k},\mathbf{p}) \mathcal{I}_{ss}\zeta_{\mathbf{p}} \zeta_{\mathbf{k}-\mathbf{p}}\,,
        \\
        h^{\lambda (2)}_{tt}(\eta, \mathbf{k}) &= \cirpol{\sum _{\lambda _1 , \lambda _2} }\frac{4}{k^3\eta} \int \frac{d^3\mathbf{p}}{(2\pi)^{\frac{3}{2}}} \left (  q^{\lambda , \lambda _1 , \lambda _2}_{tt,1}(\mathbf{k},\mathbf{p})\mathcal{I}_{tt,1} + q^{\lambda , \lambda _1 , \lambda _2}_{tt,2}(\mathbf{k},\mathbf{p}) \mathcal{I}_{tt,2}\ \right )h_{\mathbf{p}}^{\lambda _1} h_{\mathbf{k}-\mathbf{p}}^{\lambda _2}\,,
        \\
         h^{\lambda (2)}_{st}(\eta, \mathbf{k}) &=\cirpol{\sum _{\lambda _3} } \frac{8}{3k^3\eta} \int \frac{d^3\mathbf{p}}{(2\pi)^{\frac{3}{2}}} q^{\lambda , \lambda _3}_{st}(\mathbf{k},\mathbf{p}) \mathcal{I}_{st} h_{\mathbf{p}}^{\lambda _3} \zeta_{\mathbf{k}-\mathbf{p}}\,.
    \end{align}
\end{subequations}
The tensor-tensor contribution has two terms due to the nature of the source term. For the scalar-scalar and scalar-tensor we can find a common factor to factorise, $\cirpol{\left ( q^{ab}_{\lambda}(\mathbf{k}) \right ) ^*}  p_{a}p_{b}$ and $\cirpol{\left ( q^{ab}_{\lambda}(\mathbf{k}) \right ) ^*}q^{\lambda _3}_{ab}(\mathbf{p})$ respectively, this is not possible for the tensor-tensor term. In the above we introduced the kernels, $\mathcal{I} \equiv (x,p,|\mathbf{k}-\mathbf{p}|)$, defined as:
\begin{subequations}
    \begin{align}
    \refrep{\mathcal{I}_{ss}}(x,p,|\mathbf{k}-\mathbf{p}|) &= \int ^x _0 d\overline{x} f^{ss}(c_s\overline{\eta}p,c_s\overline{\eta}|\mathbf{k}-\mathbf{p}|)\overline{x}\sin(x-\overline{x})\,, \\
    \mathcal{I}_{tt,1}(x,p,|\mathbf{k}-\mathbf{p}|) &=\int ^x _0 d\overline{x} f_1^{tt}(\overline{\eta}p,\overline{\eta}|\mathbf{k}-\mathbf{p}|)\overline{x}\sin(x-\overline{x})\,,\\
    \mathcal{I}_{tt,2}(x,p,|\mathbf{k}-\mathbf{p}|) &= \int ^x _0 d\overline{x} f_2^{tt}(\overline{\eta}p,\overline{\eta}|\mathbf{k}-\mathbf{p}|)\overline{x}\sin(x-\overline{x})\,,\\
    \mathcal{I}_{st}(x,p,|\mathbf{k}-\mathbf{p}|) &= \int ^x _0 d\overline{x} f^{st}(\overline{\eta}p,c_s\overline{\eta}|\mathbf{k}-\mathbf{p}|)\overline{x}\sin(x-\overline{x})\,,
    \end{align}
\end{subequations}
where $\overline{x}=k\overline{\eta}$. We note that the pure scalar Kernel has been defined in this particular way in order to use the results from Ref.~\cite{Kohri:2018awv}. 

Using Wick's theorem and \refrep{assuming linear perturbations are Gaussian}, the two point function reduces to the sum of the three individual contributions\footnote{
We do not include contributions from third order terms, such as $\langle h^{\lambda}(\eta, \mathbf{k})h^{\lambda^{\prime}(3)}(\eta, \mathbf{k^{\prime}})\rangle$. 
We 
thank G. Dom\`{e}nech for pointing this out. \refrep{See Ref.~\cite{Chen:2022dah} for more details.}}  (scalar-scalar, tensor-tensor and scalar-tensor):
\begin{equation}\label{2pointSOGWs}
\begin{split}
\langle h^{\lambda(2)}(\eta, \mathbf{k})h^{\lambda^{\prime}(2)}(\eta, \mathbf{k^{\prime}})\rangle =&  \langle h^{\lambda(2)}_{ss}(\eta, \mathbf{k})h^{\lambda^{\prime}(2)}_{ss}(\eta, \mathbf{k^{\prime}})\rangle + \langle h^{\lambda(2)}_{tt}(\eta, \mathbf{k})h^{\lambda^{\prime}(2)}_{tt}(\eta, \mathbf{k^{\prime}})\rangle \\
+ & \langle h^{\lambda(2)}_{st}(\eta, \mathbf{k})h^{\lambda^{\prime}(2)}_{st}(\eta, \mathbf{k^{\prime}})\rangle \,.
\end{split}
\end{equation}
Here we have used the fact that the two point function $\langle \mathbf{\zeta_{\mathbf{p}}}h_{\mathbf{p^{\prime}}}\rangle$ vanishes for any momenta ${\mathbf{p}}$, ${\mathbf{p^{\prime}}}$. This can be seen as a consequence of the SVT theorem: everything decouples at linear order.

As stated in the introduction, the total power spectrum $\mathcal{P}_{h^{(2)}}$ can therefore be expressed as the sum of the three separate contributions $\mathcal{P}_{ss}(\eta ,k)$, $\mathcal{P}_{tt}(\eta ,k)$ and $\mathcal{P}_{st}(\eta ,k)$, as
\begin{eqnarray} \label{2pointgeneral}
\langle h^{\lambda(2)}(\eta, \mathbf{k})h^{\lambda^{\prime}(2)}(\eta, \mathbf{k^{\prime}})\rangle &=& \delta ^{(3)}(\mathbf{k} + \mathbf{k^{\prime}}) \delta ^{\lambda \lambda ^{\prime}} \frac{2\pi ^2}{k^3} \{ \mathcal{P}_{ss}(\eta ,k) + \mathcal{P}_{tt}(\eta ,k) + \mathcal{P}_{st}(\eta ,k) \} \,.
\end{eqnarray}
It is useful to introduce the variables:
\begin{equation} \label{uvcoordinates}
    v=\frac{p}{k}, \quad u=\frac{|\mathbf{k}-\mathbf{p}|}{k} \,,
\end{equation}
so that the integral over the wave-number $d^3\mathbf{p}= dp d\theta _p d\phi _p p^2 \sin{\theta _p}$ can be written as:
\begin{equation}
    \int d^3\mathbf{p}=k^3 \int_0^{\infty}dv \int_{|v-1|}^{1+v}du \text{ }  uv  \int_0^{2\pi}d\phi _p \,.
\end{equation}
The reader can refer to App.~\ref{Appendixsmallf} to see how the functions presented in Eq.~(\ref{sourcetermsfourier}) are expressed in the new coordinate system.

In the following, we calculate each contribution to the power spectrum of SOGWs and assume no parity violation in the primordial tensor power spectrum.
\subsection{Scalar-scalar power spectrum}
In this section we present the results for the pure scalar contribution to the final power spectrum. This has derived in Refs.~\cite{Domenech:2021ztg,Kohri:2018awv,Espinosa:2018eve} for example, which the reader can refer to for more details. Below, we recall the main steps. We need to consider: $ \langle h^{\lambda(2)}_{ss}(\eta, \mathbf{k})h^{\lambda^{\prime}(2)}_{ss}(\eta, \mathbf{k^{\prime}})\rangle$, which is given by:
\begin{eqnarray} \label{SStwopoint}
     \langle h^{\lambda(2)}_{ss}(\eta, \mathbf{k})h^{\lambda^{\prime}(2)}_{ss}(\eta, \mathbf{k^{\prime}})\rangle &=&\left ( \frac{16}{9\eta} \right )^2 \frac{1}{k^3k^{\prime 3}} \int \frac{d^3\mathbf{p}}{(2\pi)^{\frac{3}{2}}} \int \frac{d^3\mathbf{p^{\prime}}}{(2\pi)^{\frac{3}{2}}}q^{\lambda}_{ss}(\mathbf{k},\mathbf{p}) q^{\lambda ^{\prime}}_{ss}(\mathbf{k}^{\prime},\mathbf{p}^{\prime}) \nonumber \\
     &&\times I^{ss}(x,v,u)I^{ss}(x^{\prime},u^{\prime},v^{\prime})\langle \zeta _{\mathbf{p}}\zeta _{\mathbf{k-p}}\zeta _{\mathbf{p^{\prime}}}\zeta _{\mathbf{k^{\prime}-p^{\prime}}}\rangle\,.
\end{eqnarray}
The scalar four point function can be broken down into two-point correlations (Wick's theorem):
\begin{eqnarray}
    \langle \zeta _{\mathbf{p}}\zeta _{\mathbf{k-p}}\zeta _{\mathbf{p^{\prime}}}\zeta _{\mathbf{k^{\prime}-p^{\prime}}}\rangle &=& \langle \zeta _{\mathbf{p}}\zeta _{\mathbf{p^{\prime}}}\rangle \langle\zeta _{\mathbf{k-p}}\zeta _{\mathbf{k^{\prime}-p^{\prime}}}\rangle + \langle \zeta _{\mathbf{p}}\zeta _{\mathbf{k^{\prime}-p^{\prime}}}\rangle \langle\zeta _{\mathbf{k-p}}\zeta _{\mathbf{p^{\prime}}}\rangle \nonumber \\
    &=&\delta (\mathbf{k}+ \mathbf{k^{\prime}})  \left (\delta (\mathbf{p}+\mathbf{p^{\prime}}) + \delta (\mathbf{p}+\mathbf{k^{\prime}}-\mathbf{p^{\prime}}) \right ) \nonumber \\
    &\times & \frac{2\pi^2}{p^3}\frac{2\pi^2}{|\mathbf{k}-\mathbf{p}|^3} \mathcal{P}_{\zeta}(p)\mathcal{P}_{\zeta} (|\mathbf{k}-\mathbf{p}|)\,.
\end{eqnarray}
Inserting the above into Eq.~(\ref{SStwopoint}) \cirpol{and integrating over $\mathbf{p^{\prime}}$} we find:
\begin{eqnarray}
    \langle h^{\lambda(2)}_{ss}(\eta, \mathbf{k})h^{\lambda^{\prime}(2)}_{ss}(\eta, \mathbf{k^{\prime}})\rangle &=& \delta ^{(3)} (\mathbf{k} + \mathbf{k^{\prime}}) \delta ^{\lambda \lambda ^{\prime}} \frac{2\pi ^2}{k^3} \int_0^{\infty}du \int_{|1-u|}^{u+1}dv \frac{1}{x^2}\frac{64}{81}\mathcal{P}_{h}(kv) \mathcal{P}_{\zeta}(ku)\nonumber \\
    &\times&\frac{v^2}{u^2}\left (1-\frac{(1+v^2-u^2)^2}{4v^2} \right )^2\mathcal{I}_{ss}^2(x,v,u) \,,
\end{eqnarray}
and we can extract the power spectrum:
\begin{equation}\label{powerspectras}
    \mathcal{P}_{ss}(\eta ,k) = \frac{64}{81} \int_0^{\infty}dv \int_{|v-1|}^{1+v}du \frac{1}{x^2} \frac{v^2}{u^2}\bigg (1-\frac{(1+v^2-u^2)^2}{4v^2} \bigg )^2\mathcal{I}_{ss}^2(x,v,u) \mathcal{P}_{\zeta}(ku)\mathcal{P}_{\zeta}(kv)\,. 
\end{equation}
\subsection{Tensor-tensor power spectrum}
In this section we are interested in extracting the pure tensor power spectrum contribution, corresponding to $\langle h^{\lambda(2)}_{tt}(\eta, \mathbf{k})h^{\lambda^{\prime}(2)}_{tt}(\eta, \mathbf{k^{\prime}})\rangle$ in Eq.~(\ref{2pointSOGWs}):
\begin{eqnarray}\label{twopointtensor}
    \langle h^{\lambda(2)}_{tt}(\eta, \mathbf{k})h^{\lambda^{\prime}(2)}_{tt}(\eta, \mathbf{k^{\prime}})\rangle &=& \cirpol{\sum _{\lambda _1 , \lambda _2 , \lambda _1 ^{\prime}, \lambda _2 ^{\prime}} } \left (\frac{4}{\eta} \right )^2 \frac{1}{k^3k^{\prime 3}} \int \frac{d^3\mathbf{p}}{(2\pi)^{\frac{3}{2}}} \int \frac{d^3\mathbf{p^{\prime}}}{(2\pi)^{\frac{3}{2}}} \bigg ( q^{\lambda , \lambda _1 , \lambda _2}_{tt,1}(\mathbf{k},\mathbf{p})\mathcal{I}_{tt,1}(x,v,u)\nonumber \\
    &&+q^{\lambda , \lambda _1 , \lambda _2}_{tt,2}(\mathbf{k},\mathbf{p})\mathcal{I}_{tt,2}(x,v,u) \bigg ) \bigg ( q^{\lambda ^{\prime} , \lambda _1 ^{\prime}, \lambda _2^{\prime}}_{tt,1}(\mathbf{k}^{\prime},\mathbf{p}^{\prime})\mathcal{I}_{tt,1}(x^{\prime},u^{\prime},v^{\prime})\nonumber \\
    &&+q^{\lambda ^{\prime} , \lambda _1 ^{\prime}, \lambda _2^{\prime}}_{tt,2}(\mathbf{k}^{\prime},\mathbf{p}^{\prime})\mathcal{I}_{tt,2}(x^{\prime},u^{\prime},v^{\prime}) \bigg )\langle h ^{\lambda _1}_{\mathbf{p}}h ^{\lambda _2}_{\mathbf{k-p}}h ^{\lambda _1^{\prime}}_{\mathbf{p^{\prime}}}h ^{\lambda _2^{\prime}}_{\mathbf{p^{\prime}-k^{\prime}}}\rangle\,.
\end{eqnarray}
\cirpol{
The tensor four-point correlation function is given by:
\begin{eqnarray}\label{ttcorrelation}
    \langle h ^{\lambda _1}_{\mathbf{p}}h ^{\lambda _2}_{\mathbf{k-p}}h ^{\lambda _1^{\prime}}_{\mathbf{p^{\prime}}}h ^{\lambda _2^{\prime}}_{\mathbf{p^{\prime}-k^{\prime}}}\rangle  &=& \langle h ^{\lambda _1}_{\mathbf{p}}h ^{\lambda _1^{\prime}}_{\mathbf{p^{\prime}}} \rangle \langle h ^{\lambda _2}_{\mathbf{k-p}}h ^{\lambda _2^{\prime}}_{\mathbf{p^{\prime}-k^{\prime}}}\rangle + \langle h ^{\lambda _1}_{\mathbf{p}}h ^{\lambda _2^{\prime}}_{\mathbf{p^{\prime}-k^{\prime}}} \rangle \langle h ^{\lambda _2}_{\mathbf{k-p}}h ^{\lambda _1^{\prime}}_{\mathbf{p^{\prime}}}\rangle \nonumber \\
    &=&\delta (\mathbf{k}+ \mathbf{k^{\prime}}) \delta (\mathbf{p}+\mathbf{p^{\prime}}) \left (\delta ^{\lambda_1 \lambda_1^{\prime}}\delta ^{\lambda_2 \lambda_2^{\prime}} +\delta ^{\lambda_1 \lambda_2^{\prime}}\delta ^{\lambda_2 \lambda_1^{\prime}} \right ) \nonumber \\
    &\times& \frac{2\pi^2}{p^3}\frac{2\pi^2}{|\mathbf{k}-\mathbf{p}|^3} \mathcal{P}_{h}(p)\mathcal{P}_{h} (|\mathbf{k}-\mathbf{p}|)\,.
\end{eqnarray}
}
Here $\delta^{\lambda_1 \lambda_1^{\prime}}$ is the Kronecker delta which ensures the correct polarizations are selected. In App.~\ref{symmetryproof} we show that the integral over $\mathbf{p}$ is invariant under $\mathbf{p}\rightarrow \mathbf{k}-\mathbf{p}$. This implies that when carrying out the integral over $\mathbf{p}^{\prime}$ the contribution from integrating over $\delta (\mathbf{p}+\mathbf{p^{\prime}})$ or $\delta (\mathbf{p}+\mathbf{k^{\prime}}-\mathbf{p^{\prime}})$ is identical, \cirpol{hence the factor of $\delta (\mathbf{k}+ \mathbf{k^{\prime}}) \delta (\mathbf{p}+\mathbf{p^{\prime}})$ in the second line of the above equation. The polarization functions are evaluated in App.~\ref{azimuthalstuff} and for clarity we show how we construct the convoluted polarization tensors in spherical coordinates in App.~\ref{setuptt}. Integrating over the azimuthal plane and switching to coordinates $v$ and $u$ leads to}:\cirpol{
\begin{eqnarray} \label{tensorpowerspec}
    &&\langle  h^{\lambda(2)}_{tt}(\eta, \mathbf{k})h^{\lambda^{\prime}(2)}_{tt}(\eta, \mathbf{k^{\prime}})\rangle =  \delta ^{(3)} (\mathbf{k} + \mathbf{k^{\prime}}) \delta ^{\lambda \lambda ^{\prime}} \frac{2\pi ^2}{k^3}   \int_0^{\infty}dv \int_{|v-1|}^{1+v}du \frac{1}{x^2} \frac{16}{k^4 u^6v^6} \nonumber \\
    &\times &\bigg ( \frac{k^4}{32768} A(v,u)E^2(v,u)\mathcal{I}^2_{tt,1}(x,v,u) +\frac{k^4}{32768}D(v,u)H^2(v,u) \mathcal{I}^2_{tt,1}(x,v,u) \nonumber \\
    & -&\frac{k^2}{4096}A(v,u)E(v,u) \mathcal{I}_{tt,1}(x,v,u) \mathcal{I}_{tt,2}(x,v,u)-\frac{k^2}{4096}D(v,u)H(v,u) \mathcal{I}_{tt,1}(x,v,u) \mathcal{I}_{tt,2}(x,v,u)\nonumber \\
    &+&\frac{1}{2028}A(v,u)\mathcal{I}^2_{tt,2}(x,v,u) +\frac{1}{2028}D(v,u)\mathcal{I}^2_{tt,2}(x,v,u) + \frac{k^4}{65536} B(v,u)F^2(v,u)\mathcal{I}^2_{tt,1}(x,v,u) \nonumber \\
    &+&\frac{k^4}{65536} C(v,u)G^2(v,u)\mathcal{I}^2_{tt,1}(x,v,u) + \frac{k^4}{32768} I(v,u)G(v,u)F(v,u)\mathcal{I}^2_{tt,1}(x,v,u)  -\frac{k^2}{8192}\nonumber \\
    &\times &B(v,u)F(v,u) \mathcal{I}_{tt,1}(x,v,u) \mathcal{I}_{tt,2}(x,v,u) -\frac{k^2}{8192}C(v,u)G(v,u) \mathcal{I}_{tt,1}(x,v,u) \mathcal{I}_{tt,2}(x,v,u)\nonumber \\
    & -&\frac{k^2}{8192}I(v,u)F(v,u) \mathcal{I}_{tt,1}(x,v,u) \mathcal{I}_{tt,2}(x,v,u)  -\frac{k^2}{8192}I(v,u)G(v,u) \mathcal{I}_{tt,1}(x,v,u) \mathcal{I}_{tt,2}(x,v,u)\nonumber  \\
    &+&\frac{1}{4056}B(v,u)\mathcal{I}^2_{tt,2}(x,v,u) +\frac{1}{4056}C(v,u)\mathcal{I}^2_{tt,2}(x,v,u)  +\frac{1}{2028}I(v,u)\mathcal{I}^2_{tt,2}(x,v,u)   \bigg ) \nonumber \\
    &\times &\mathcal{P}_{h}(ku)\mathcal{P}_{h}(kv) \, , 
\end{eqnarray}}
\cirpol{where the functions $A(v,u)$, $B(v,u)$, $C(v,u)$, $D(v,u)$, $E(v,u)$, $F(v,u)$, $G(v,u)$, $H(v,u)$ and $I(v,u)$ that appear in front of the kernels are defined in App.~\ref{azimuthalstuff}.} Similarly to Ref.~\cite{Espinosa:2018eve} we split our Kernels as:
\begin{subequations}
\begin{align}
\begin{split}
    \mathcal{I}_{tt,1}^2(x,v,u) &= \cos ^2(x)[\mathcal{I}^{tt}_{c,1}(x,v,u)]^2 + \sin ^2(x)[\mathcal{I}^{tt}_{s,1}(x,v,u)]^2 \\
    &+ \sin (2x)\mathcal{I}^{tt}_{c,1}(x,v,u)\mathcal{I}^{tt}_{s,1}(x,v,u) \,,
\end{split}
\\
\begin{split}
    \mathcal{I}_{tt,2}^2(x,v,u) &= \cos ^2(x)[\mathcal{I}^{tt}_{c,2}(x,v,u)]^2 + \sin ^2(x)[\mathcal{I}^{tt}_{s,2}(x,v,u)]^2 \\
    &+ \sin (2x)\mathcal{I}^{tt}_{c,2}(x,v,u)\mathcal{I}^{tt}_{s,2}(x,v,u)\,,
\end{split}
\\
\begin{split}
    \mathcal{I}_{tt,1}(x,v,u)\mathcal{I}_{tt,2}(x,v,u) &= \cos ^2(x)\mathcal{I}^{tt}_{c,1}(x,v,u)\mathcal{I}^{tt}_{c,2}(x,v,u) + \sin ^2(x)\mathcal{I}^{tt}_{s,1}(x,v,u)\mathcal{I}^{tt}_{s,2}(x,v,u) \\
        &+\cos{x}\sin{x} \mathcal{I}^{tt}_{c,1}(x,v,u)\mathcal{I}^{tt}_{s,2}(x,v,u) +\cos{x}\sin{x} \mathcal{I}^{tt}_{s,1}(x,v,u)\mathcal{I}^{tt}_{c,2}(x,v,u)
\end{split} 
\end{align}
\end{subequations}
where 
\begin{subequations} \label{ttkernel}
\begin{align}
 &\mathcal{I}_{c,1}^{tt} (x,v,u) = \int _0^{x} d\overline{x} \sin(-\overline{x})\overline{x} f^{tt}_1(\overline{x},v,u)\,, \\
&\mathcal{I}_{s,1}^{tt} (x,v,u) = \int _0^{x} d\overline{x} \cos(\overline{x})\overline{x} f^{tt}_1(\overline{x},v,u)\,, \\
         &\mathcal{I}_{c,2}^{tt} (x,v,u) = \int _0^{x} d\overline{x} \sin(-\overline{x}) \overline{x}f^{tt}_2(\overline{x},v,u)\,, \\
        &\mathcal{I}_{s,2}^{tt} (x,v,u) = \int _0^{x} d\overline{x} \cos(\overline{x}) \overline{x}f^{tt}_2(\overline{x},v,u)\,.
\end{align}
\end{subequations}
By comparing Eq.~(\ref{2pointgeneral}) to Eq.~(\ref{tensorpowerspec}) it is straightforward to see what contributes to the final power spectrum:
\cirpol{
\begin{align}
    \begin{split}
        &\mathcal{P}_{tt}(\eta ,k)= \frac{16}{x^2} \int_0^{\infty}dv \int_{|v-1|}^{1+v}du \text{ }   \frac{1}{k^4u^6v^6}  \bigg ( \frac{k^4}{65536} \big (2A(v,u)E^2(v,u) + 2D(v,u)H^2(v,u) \\
        &+ B(v,u)F^2(v,u) + C(v,u)G^2(v,u) + 2I(v,u)G(v,u)F(v,u) \big ) \mathcal{I}^2_{tt,1}(x,v,u)    \\
        & -\frac{k^2}{8192}\big (2A(v,u)E(v,u)+2D(v,u)H(v,u)+B(v,u)F(v,u)+C(v,u)G(v,u)\\
        &+I(v,u)F(v,u)+I(v,u)G(v,u) \big ) \mathcal{I}_{tt,1}(x,v,u) \mathcal{I}_{tt,2}(x,v,u)+\frac{1}{4056}\big (2A(v,u) \\
        & +2D(v,u)+B(v,u)+C(v,u)+2I(v,u)\big )\mathcal{I}^2_{tt,2}(x,v,u) \bigg ) \mathcal{P}_{h}(kv) \mathcal{P}_{h}(ku) \, .
    \end{split}
\end{align}}

\subsection{Scalar-tensor power spectrum}
In this section we show how to extract the mixed-scalar power spectrum contribution, corresponding to $\langle h^{\lambda(2)}_{st}(\eta, \mathbf{k})h^{\lambda^{\prime}(2)}_{st}(\eta, \mathbf{k^{\prime}})\rangle$ in Eq.~(\ref{2pointSOGWs}):
\begin{eqnarray}
     \langle h^{\lambda(2)}_{st}(\eta, \mathbf{k})h^{\lambda^{\prime}(2)}_{st}(\eta, \mathbf{k^{\prime}})\rangle  &=& \cirpol{ \sum _ {\lambda _3, \lambda _3 ^{\prime}}} \left ( \frac{8}{3\eta} \right )^2 \frac{1}{k^3k^{\prime 3}} \int \frac{d^3\mathbf{p}}{(2\pi)^{\frac{3}{2}}} \int \frac{d^3\mathbf{p^{\prime}}}{(2\pi)^{\frac{3}{2}}}q^{\lambda,\lambda _3}_{st}(\mathbf{k},\mathbf{p}) q^{\lambda ^{\prime},\lambda _3^{\prime}}_{st}(\mathbf{k}^{\prime},\mathbf{p}^{\prime}) \nonumber \\
     &&\times \mathcal{I}_{st}(x,v,u)\mathcal{I}_{st}(x^{\prime},u^{\prime},v^{\prime})  \langle h^{\lambda _3}_{\mathbf{p}}\zeta _{\mathbf{k-p}}h^{\lambda _3^{\prime}}_{\mathbf{p^{\prime}}}\zeta _{\mathbf{k^{\prime}-p^{\prime}}}\rangle \,.
\end{eqnarray}
The four-point correlation function is given by:
\begin{eqnarray}
    \langle h^{\lambda _3}_{\mathbf{p}}\zeta _{\mathbf{k-p}}h^{\lambda _3^{\prime}}_{\mathbf{p^{\prime}}}\zeta _{\mathbf{k^{\prime}-p^{\prime}}}\rangle &=& \langle h^{\lambda _3}_{\mathbf{p}}h^{\lambda _3^{\prime}}_{\mathbf{p^{\prime}}}\rangle \langle \zeta _{\mathbf{k-p}}\zeta \mathbf{ _{k^{\prime}-p^{\prime}}}\rangle \nonumber \\
    &=&\delta (\mathbf{k}+ \mathbf{k^{\prime}}) \delta (\mathbf{p}+\mathbf{p^{\prime}})\delta ^{\lambda_3 \lambda_3^{\prime}} \frac{2\pi^2}{p^3}\frac{2\pi^2}{|\mathbf{k}-\mathbf{p}|^3} \mathcal{P}_{h}(p)\mathcal{P}_{\zeta} (|\mathbf{k}-\mathbf{p}|)\,.
\end{eqnarray}
Following the same steps outlined in the previous section:\cirpol{
\begin{eqnarray}
    \langle h^{\lambda(2)}_{st}(\eta, \mathbf{k})h^{\lambda^{\prime}(2)}_{st}(\eta, \mathbf{k^{\prime}})\rangle &=& \delta ^{(3)} (\mathbf{k} + \mathbf{k^{\prime}}) \delta ^{\lambda \lambda ^{\prime}} \frac{2\pi ^2}{k^3} \int_0^{\infty}dv \int_{|1-v|}^{v+1}du \frac{1}{36x^2}\mathcal{P}_{h}(kv) \mathcal{P}_{\zeta}(ku)\nonumber \\
    &\times&\frac{1}{k^4u^2v^6}\left ( (u^2-(1+v)^2)^4 + (u^2-(-1+v)^2)^4 \right )\mathcal{I}_{st}^2(x,v,u)  \,. \nonumber \\
\end{eqnarray}}
Similarly, the kernels are split up as:
\begin{equation}
\begin{split}
    \mathcal{I}_{st}^2(x,v,u) &= \cos ^2(x)[\mathcal{I}^{st}_{c}(x,v,u)]^2 + \sin ^2(x)[\mathcal{I}^{st}_{s}(x,v,u)]^2 \\
    &+ \sin (2x)\mathcal{I}^{st}_{c}(x,v,u)\mathcal{I}^{st}_{s}(x,v,u) \,,
\end{split}
\end{equation}
where 
\begin{subequations} \label{stkernel}
\begin{align}
 &\mathcal{I}_{c}^{st} (x,v,u) = \int _0^{x} d\overline{x} \sin(-\overline{x}) \overline{x}f^{st}(\overline{x},v,u)\,, \\
        &\mathcal{I}_{s}^{st} (x,v,u) = \int _0^{x} d\overline{x} \cos(\overline{x}) \overline{x}f^{st}(\overline{x},v,u) \,.
\end{align}
\end{subequations}
Hence we have:
\cirpol{
\begin{eqnarray}\label{powerpectrumst}
    \mathcal{P}_{st}(\eta ,k) &=& \frac{1}{x^2}\frac{1}{36} \int_0^{\infty}dv \int_{|v-1|}^{1+v}du  \frac{1}{k^4u^2v^6}\left ( (u^2-(1+v)^2)^4 + (u^2-(-1+v)^2)^4 \right )\mathcal{I}_{st}^2(x,v,u) \nonumber \\ 
    &\times & \mathcal{P}_{h}(kv) \mathcal{P}_{\zeta}(ku) \,.
\end{eqnarray}
}

\section{Results: peaked power spectra}\label{section4}
In this section we present our results: expressions for the current spectral density of SOGWs. Furthermore, we show the resulting spectral density for peaked power spectra sources, comment on the behaviour of the individual contributions as well as the total contribution. We also make some comments on detectability.
\subsection{Current energy density of gravitational waves}
\refrep{In order to relate our computations to future gravitational wave detectors, we use the spectral density} \cite{Caprini:2018mtu}, related to the power spectrum by
\begin{equation}\label{spectraldensity}
    \Omega _{GW}(\eta, k) = \frac{1}{6} \left ( \frac{k}{\mathcal{H(\eta)}} \right )^2 \overline{\mathcal{P}_{h^{(2)}}(\eta, k)} \,,
\end{equation}
where the overline here denotes an oscillation average over a few periods.
Since these waves are generated in an early radiation dominated universe, we have to take into account how their density will change as the universe evolves. The density of gravitational waves scales like radiation and therefore it is possible to relate the spectral density at the time of creation, $\Omega _{GW}(k)$, to the present the spectral density, $\Omega _{GW}(\eta _0 , k)$, which takes into account how the universe becomes matter dominated after the matter-radiation equality. For a detailed explanation the reader can refer to Ref.~\cite{Domenech:2021ztg}:
\begin{equation} \label{GWdensity}
    \Omega _{GW}(\eta _0,k)h^2 =1.62\times10^{-5}\times \Omega _{GW}(\eta, k) =1.62\times10^{-5}\times \frac{1}{6} \frac{k^2}{\mathcal{H}(\eta )^2} \overline{\mathcal{P}_{h^{(2)}}(\eta , k)} \,,
\end{equation}
where $h$ is the dimensionless reduced Hubble constant to take into account the uncertainty in its value. Using $\mathcal{H}(\eta )=1/\eta $ we define:
\begin{eqnarray}
    \Omega _{GW}(\eta _0,k)h^2 = 1.62\times10^{-5}\times  \left \{ \Omega _{ss}(\eta,k) + \Omega _{tt}(\eta,k) + \Omega _{st}(\eta,k) \right \}\,,
\end{eqnarray}
where $\Omega _{ss}(\eta,k)$, $\Omega _{tt}(\eta,k)$ and $\Omega _{st}(\eta,k)$ are respectively the spectral density from the induced scalar-scalar, tensor-tensor and scalar-tensor at the time of creation. Below we present the corresponding expressions and their kernels which are the main results of our paper. To see the details behind our kernel computations see App.~\ref{kerneleval}. We note that the spectral density depends on the time-averaged power spectrum. The time dependence in our expressions is contained in the kernels, via the variable $x=k\eta$, hence the following results are used to derive the final expressions:
\begin{equation*}
    \overline{\sin x}=\overline{\cos x}=0, \quad \overline{\sin ^2 x}=\overline{\cos ^2 x}=\frac{1}{2}\,.
\end{equation*}
\subsubsection{Spectral density of the scalar-scalar contribution}\label{resultssecss}
Inserting the power spectrum Eq.~(\ref{powerspectras}) into Eq.~(\ref{spectraldensity}), we arrive to an expression for the spectral density from the scalar-scalar contribution
\begin{equation} \label{densityss}
    \Omega _{ss}(\eta,k) =\frac{32}{243}  \int_0^{\infty}dv \int_{|v-1|}^{1+v}du \bigg (\frac{4v^2-(1+v^2-u^2)^2}{4uv} \bigg )^2\overline{\mathcal{I}_{ss}^2(v,u)} \mathcal{P}_{\zeta}(ku)\mathcal{P}_{\zeta}(kv)\,,
\end{equation}
with
\begin{eqnarray}\label{SSfullkernel}
    \overline{\mathcal{I}_{ss}^2(v,u)} &=& 2 \left ( \frac{3(u^2+v^2-3)}{4u^3v^3}\right )^2 \bigg [\left ( -4uv+(u^2+v^2-3) \log \bigg |\frac{3-(u+v)^2}{3-(u-v)^2} \bigg |\right )^2 \nonumber \\
    &&+\pi ^2 (u^2+v^2-3)^2 \Theta (v+u-\sqrt{3})  \bigg]\,.
\end{eqnarray}
The expression for the Kernel has been \refrep{taken} from Ref.~\cite{Kohri:2018awv}.  
\subsubsection{\refrep{Spectral density of the tensor-tensor contribution}}\label{resultssectt}
Similarly, the tensor-tensor spectral density is given by:
\cirpol{
\begin{align} \label{densitytt}
    \begin{split}
        &\Omega _{tt}(\eta,k)= \frac{8}{3} \int_0^{\infty}dv \int_{|v-1|}^{1+v}du \text{ }   \frac{1}{k^4u^6v^6}  \bigg ( \frac{k^4}{65536} \big (2A(v,u)E^2(v,u) + 2D(v,u)H^2(v,u) \\
        &+ B(v,u)F^2(v,u) + C(v,u)G^2(v,u) + 2I(v,u)G(v,u)F(v,u) \big ) \mathcal{I}^2_{tt,1}(v,u)    \\
        & -\frac{k^2}{8192}\big (2A(v,u)E(v,u)+2D(v,u)H(v,u)+B(v,u)F(v,u)+C(v,u)G(v,u)\\
        &+I(v,u)F(v,u)+I(v,u)G(v,u) \big ) \mathcal{I}_{tt,1}(v,u) \mathcal{I}_{tt,2}(k,v,u)+\frac{1}{4056}\big (2A(v,u) \\
        & +2D(v,u)+B(v,u)+C(v,u)+2I(v,u)\big )\mathcal{I}^2_{tt,2}(k,v,u) \bigg ) \mathcal{P}_{h}(kv) \mathcal{P}_{h}(ku) \, ,
    \end{split}
\end{align}}
\cirpol{where 
\begin{subequations}
\begin{align}
        A(v,u)&= (1+u-v)^2(1-u+v)^2(-1+u+v)^2(1+u+v)^6 \, ,
        \\
        B(v,u)&= (1+u-v)^2(1-u+v)^6(-1+u+v)^2(1+u+v)^2 \, ,
        \\
        C(v,u)&= (1+u-v)^6(1-u+v)^2(-1+u+v)^2(1+u+v)^2 \, ,
        \\
        D(v,u)&= (1+u-v)^2(1-u+v)^2(-1+u+v)^6(1+u+v)^2 \, ,
        \\
        E(v,u)&= 9u^2-2u(v-1)+(v+3)(v-1)\, ,
        \\
        F(v,u)&= 9u^2+2u(v-1)+(v+3)(v-1)\, ,
        \\
        G(v,u)&= 9u^2+2u(v+1)+(v-3)(v+1)\, ,
        \\
        H(v,u)&= 9u^2-2u(v+1)+(v-3)(v+1)\, ,
        \\
        I(v,u)&= \left ( (u-v)^2-1 \right )^4 (-1+u+v)^2(1+u+v)^2 \, ,
\end{align}
\end{subequations}
}
and the time-averaged kernels are shown below:
\begin{subequations}\label{TTfullkernel}
\begin{align}
&\overline{\mathcal{I}_{tt,1}^2(v,u)} = \frac{1}{32u^2v^2}\left(\pi ^2 \Theta (u+v-1) + \log \bigg | \frac{1-(u+v)^2}{1-(u-v)^2} \bigg |^2 \right) \,,\\
&\overline{\mathcal{I}_{tt,2}^2(k,v,u)} = \frac{k^4}{128u^2v^2}\left ( \pi ^2(u^2+v^2-1)^2\Theta (u+v-1) + \left (\log \bigg | \frac{1-(u+v)^2}{1-(u-v)^2} \bigg |-4vu \right )^2 \right ) \,,\\
&\overline{\mathcal{I}_{tt,1}(v,u)\mathcal{I}_{tt,2}(k,v,u)} = \frac{k^2}{64u^2v^2}\bigg (\pi ^2(u^2+v^2-1)\Theta (u+v-1) + \log \bigg | \frac{1-(u+v)^2}{1-(u-v)^2} \bigg | \bigg (-4vu\nonumber \\
& +\log \bigg | \frac{1-(u+v)^2}{1-(u-v)^2} \bigg |\bigg ) \bigg )\,.
\end{align}
\end{subequations}
\refrep{The spectral density expression for the tensor-tensor induced waves differs from Ref.~\cite{Yu:2023lmo}, although we have the same source terms. This is probably due to how the convoluted polarization tensors were constructed.}
\subsubsection{Spectral density of the scalar-tensor contribution}\label{resultssecst}
Finally, the spectral density of scalar-tensor induced gravitational waves reads:
\cirpol{
\begin{eqnarray}\label{densityst}
   \Omega _{st}(\eta,k) &=& \frac{1}{216} \int_0^{\infty}dv \int_{|v-1|}^{1+v}du  \frac{1}{k^4u^2v^6}\left ( (u^2-(1+v)^2)^4 + (u^2-(-1+v)^2)^4 \right )\mathcal{I}_{st}^2(x,v,u) \nonumber \\ 
    &\times & \mathcal{P}_{h}(kv) \mathcal{P}_{\zeta}(ku) \, ,
\end{eqnarray}
}
with
\begin{eqnarray}\label{STfullkernel}
\overline{\mathcal{I}_{st}^2(k,v,u)} &=& \frac{ k^4}{512u^6v^2} \bigg (3\pi ^2 \left ((u^2-3(v-1)^2)(u^2-3(v+1)^2) \right )^2\Theta (v+\frac{u}{\sqrt{3}}-1) \nonumber \\ \nonumber 
&+&\bigg \{-\sqrt{3} (u^2-3(v-1)^2)(u^2-3(v+1)^2) \log \bigg | \frac{(\sqrt{3}v-u)^2-3}{(\sqrt{3}v+u)^2-3} \bigg |     \nonumber \\
    &+& 4uv\left (u^2-9v^2+9 \right )\bigg \}^2\bigg) \,.
\end{eqnarray}

\refrepencore{Interestingly, our spectral density expression for the scalar-tensor induced waves matches Ref.~\cite{Yu:2023lmo} but differs from Ref.~\cite{Bari:2023rcw}. The expansion of the metric in Ref.~\cite{Bari:2023rcw} is done exponentially as opposed to our expansion (and Ref.~\cite{Yu:2023lmo}). Hence, our second order tensor variables are not equal\footnote{\refrepencore{The relationship between our tensor variable ($h_{ij}^{(n)}$) and the tensor variable in Ref.~\cite{Bari:2023rcw} ($\gamma_{ij}^{(n)}$) is:
\begin{equation*}
    h_{ij}^{(1)} = \frac{1}{2}\gamma _{ij}^{(1)}\, , \quad  h_{ij}^{(2)} = \gamma _{ij}^{(2)} - 2\gamma _{ij}^{(1)} \Psi ^{(1)}   \, .
\end{equation*}
}}. Since the observable we calculate, the spectral density, depends on how the second order tensors are defined, ultimately we are not comparing the same quantity. We leave this comparison for future work.}

\subsection{Comments on potential divergences and resonances} \label{divsection}
The behaviour of the integrands (\ref{densityss}), (\ref{densitytt}) and (\ref{densityst}) in the IR and UV limits contains information of any possible divergences that may appear. The IR limit (long wave-length) corresponds to taking $k\rightarrow 0$, then $v\rightarrow 1/k$ and $u\rightarrow 1/k$, and all three contributions are well behaved in this limit. In fact, in the IR limit the scalar-scalar induced waves have a log-dependent slope \cite{Yuan:2019wwo} and it was shown that this is not the case for the scalar-tensor induced waves due to scalar and tensor fluctuations not having the same propagation speed \cite{Bari:2023rcw}.

In the UV limit (short wave-length), as $k\rightarrow \infty$, we have $v\rightarrow 0$ and $u\rightarrow 1$ for the pure quadratic contributions. The scalar-scalar limit is finite, whilst the tensor-tensor limit goes as \cirpol{$v^{-4}$}. \refrep{For the cross scalar-tensor term, there are two distinct limits}: $v\rightarrow 0$ and $u\rightarrow 1$, corresponding to the large wavelength for tensors, and $v\rightarrow 1$ and $u\rightarrow 0$ corresponding to the large wavelength for scalars. In the former case the integrand diverges as \cirpol{$v^{-4}$} whilst in the latter as \refrep{$u^{-4}$}. Hence we see that if the first order power spectra are given by a power law, below a certain order, there could be an enhanced contribution to the final spectral density. \refrep{Strictly speaking, in the UV limit modes become non-classical and hence a renormalisation scheme should be employed. This was first pointed out for the scalar-tensor case in Ref.~\cite{Bari:2023rcw}, where their chosen scheme can be used to mitigate the divergence.}

Furthermore, the scalar-scalar, tensor-tensor and scalar-tensor integrands contain logarithmic terms that could lead to possible divergences, provided $v$ and $u$ take particular values. By considering `momentum conservation' we can extract a range allowed values for $v$ and $u$. The wave vectors in Fourier space of the three modes form a triangle, with sides: $\mathbf{k}$, $\mathbf{p}$ and $\mathbf{k}-\mathbf{p}$, where the magnitude of the vectors satisfy the triangle inequality, which in terms of the variables $v$ and $u$ is given by: $|u-v|<1<u+v$. Hence, we see that the tensor-tensor spectral density will not enter a logarithmic resonance since the value of $v$ and $u$ needed do not satisfy the triangle inequality. On the other hand, the scalar-scalar contribution can enter into a logarithmic resonance when $v+u=\sqrt{3}$ since this corresponds to a possible configuration of the wave-vectors. Similarly, there is a possible logarithmic divergence for the scalar-tensor contribution when $v=1\pm c_su$, but this is prevented from the terms multiplying the logarithmic terms. However, in the limit $v\rightarrow 1\pm c_su$ the scalar-tensor integrand tends to $u^{-4}$ which causes a divergences in the large scalar wavelength limit. 

\refrep{For the reasons outlined above we will exclusively use  peaked power spectra as input, which we do in the following sections, by first looking at the induced waves from monochromatic power spectra and secondly using lognormal power spectra. Having a sufficiently sharp peak means that all the power in the power spectrum will be at a particular scale and hence not encompass possible scales that will lead to a divergence.} 

\subsection{Monochromatic power spectra}
It is informative to study the spectrum of SOGWs that results from having a sharp peak in the input power spectrum, modelled by a Dirac delta function. Although `not physical' (i.e.~there is no such thing as a infinitely sharp peak in nature), the resulting spectral density tells us a lot about the properties and shape of resulting spectra in general.

The monochromatic power spectrum is given by
\begin{equation}
    \mathcal{P}_{\zeta , h}(k)= \mathcal{A}_{\zeta,h}\delta \left (\log  \frac{k}{k_{\zeta,h}} \right )\,,
\end{equation}
with $\mathcal{A}_{\zeta,h}$ the amplitude of the power spectrum and $k_{\zeta,h}$ the location of the scalar and tensor peak in the primordial power spectrum, respectively. Hence we get
\begin{equation}
    \mathcal{P}_{\zeta , h}(kv)= \mathcal{A}_{\zeta,h}\delta \left (\log \frac{kv}{k_{\zeta,h}}\right )=\mathcal{A}_{\zeta,h} \frac{k_{\zeta,h}}{k} \delta (v-\frac{k_{\zeta,h}}{k})\,.
\end{equation}
The same holds for $\mathcal{P}_{\zeta , h}(ku)$, if we swap $v\leftrightarrow u$. The way we have expressed the power spectrum makes it straightforward to use the sifting property of the Dirac delta function, we just have to take the limits of integration of Eqs. (\ref{densityss}), (\ref{densitytt}) and (\ref{densityst}) into account
\begin{eqnarray}\label{krangedirac}
     \begin{cases}|1-v|<u<1+v\\ 0<v<\infty\end{cases}  =\begin{cases}|k-k_h|<k_{\zeta}<k+k_h\\ 0<k_h/k<\infty\end{cases}  =
    |k_{\zeta}-k_h|<k<k_{\zeta}+k_h \text{.}  
\end{eqnarray}
Below we study two different scenarios: first, the case where the primordial power spectrum for scalars and tensors peaks at the same scale and then the case where the two peaks are at different scales.

\subsubsection{Peaks at the same scale}
In this section, the primordial power spectrum of scalar and tensor perturbations peaks at the same scale so that $k_{\zeta}=k_h\equiv k_p$. The inequality (\ref{krangedirac}) reduces to: $0<k<2k_p$, which can be encoded by a Heaviside step function. For convenience, we define $\Tilde{k}=k/k_p$.

The analytic expression for scalar-scalar part Eq.~(\ref{densityss}) is
\begin{eqnarray}\label{DDscalarscalar}
    \Omega _{ss}(\eta ,k) &=& \frac{32}{243} \mathcal{A}_{\zeta}^2\Tilde{k}^{-2}(\frac{\Tilde{k}^2-4}{4})^2 \overline{I^{ss}(v=\Tilde{k}^{-1},u=\Tilde{k}^{-1})^2}\Theta(2k_p-k) \,.
\end{eqnarray}
\begin{figure} 
    \centering
    \includegraphics[width=0.7\textwidth]{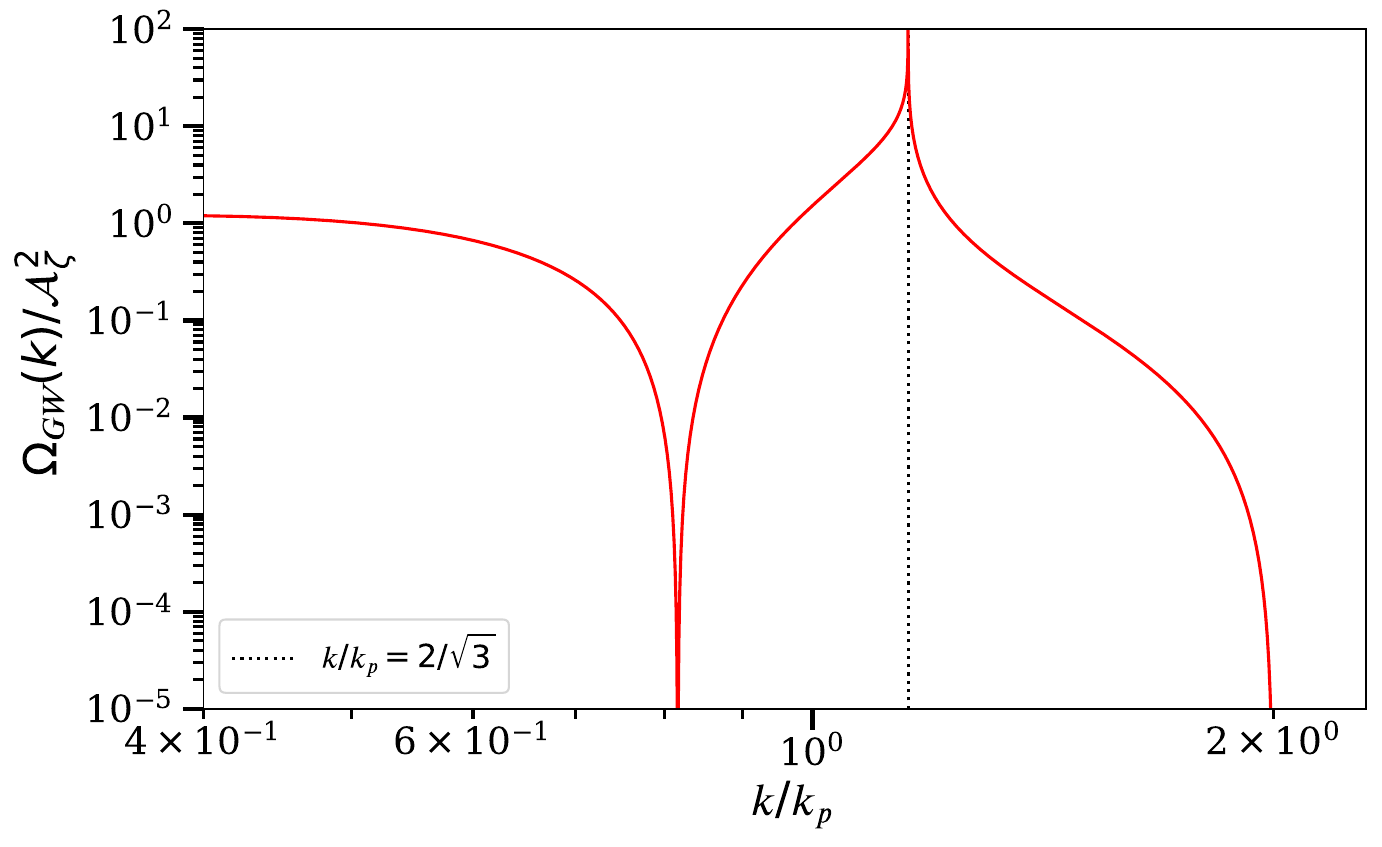}
    \caption{The spectral density $ \Omega _{ss}(k)$ of second order gravitational waves induced by scalar-scalar modes, normalised by $\mathcal{A}_{\zeta}$,  against a range of $k$ around some pivot scale $k_p$. There is a logarithmic resonance that occurs when $\Tilde{k}=2/\sqrt{3}$, represented by the black-dotted line, as expected from looking at the kernel (\ref{SSfullkernel}). After the peak, the spectral density decreases rapidly before hitting the cut-off at $k=2k_p$.}
    \label{DDss}
\end{figure}
The resulting behaviour of Eq.~(\ref{DDscalarscalar}) is shown in Fig.~\ref{DDss}. There is a sharp peak at $\Tilde{k}=2/\sqrt{3}$, or when $k=2c_sk_p$, which corresponds to $u+v=\sqrt{3}$ as expected following our discussion on resonances in the scalar-scalar SOGWs in Sect.~\ref{divsection}. Furthermore, due to momentum conservation, there is the expected sharp cutoff at $k=2k_p$.

For the Dirac peaked spectra the tensor-tensor part Eq.~(\ref{densitytt})
\cirpol{
\begin{align}\label{DDtensor}
    \begin{split}
        &\Omega _{tt}(\eta,k)= \frac{8}{3} \frac{\Tilde{k}^{10}}{k^4}\bigg (\frac{k^4(\Tilde{k}^2-4)^2}{8192\Tilde{k}^{12}}(512+1536\Tilde{k}^2-336\Tilde{k}^4-40\Tilde{k}^6+9\Tilde{k}^8)\overline{\mathcal{I}^2_{tt,1}(v=\Tilde{k}^{-1},u=\Tilde{k}^{-1})} \\
        &-\frac{k^2(\Tilde{k}^2-4)^2}{1024\Tilde{k}^{10}}(-64-136\Tilde{k}^2+10\Tilde{k}^4+3\Tilde{k}^6)\overline{\mathcal{I}_{tt,1}(v=\Tilde{k}^{-1},u=\Tilde{k}^{-1}) \mathcal{I}_{tt,2}(k,v=\Tilde{k}^{-1},u=\Tilde{k}^{-1})} \\
        &+\frac{(\Tilde{k}^2-4)^2}{4056\Tilde{k}^{8}}(64+96\Tilde{k}^2+7\Tilde{k}^4)\overline{\mathcal{I}^2_{tt,2}(k,v=\Tilde{k}^{-1},u=\Tilde{k}^{-1})}    \bigg ) \Theta(2k_p-k)\,.
    \end{split}
\end{align}
}
\begin{figure} 
    \centering
    \includegraphics[width=0.7\textwidth]{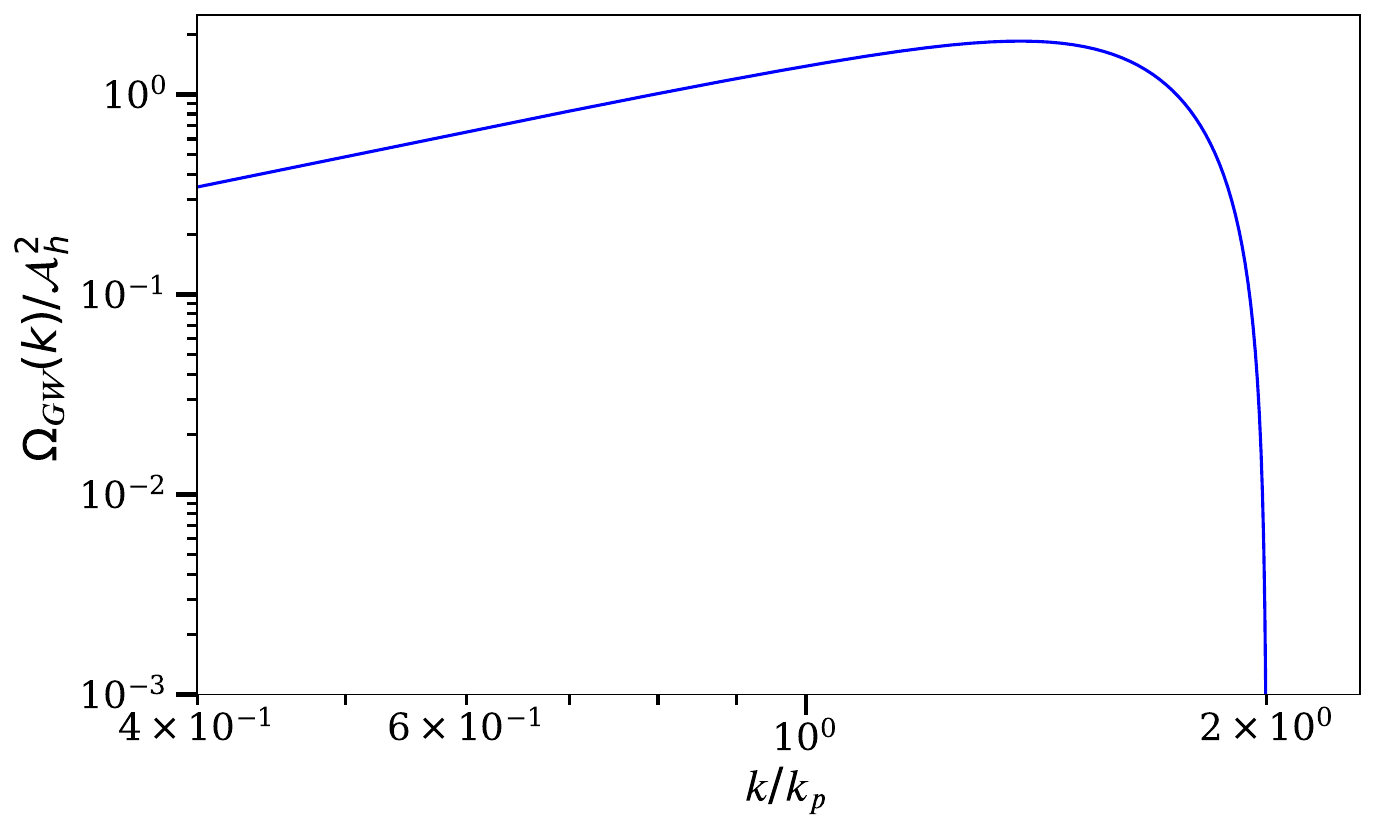}
    \caption{The spectral density $\Omega _{tt}(k)$ of second order gravitational waves induced by tensor-tensor modes, normalised by $\mathcal{A}_{h}$, against a range of $k$ around some pivot scale $k_p$. There is no logarithmic resonance since the magnitude of the wave-vectors needed for the resonance, $u+v=1$, are not allowed due to momentum conservation.}
    \label{DDtt}
\end{figure}
Equation (\ref{DDtensor}) is plotted in Fig \ref{DDtt}. We note that, as expected,  for tensor-tensor induced gravitational waves there is no resonant peak.

The scalar-tensor or cross-term part of Eq.~(\ref{densityst}) is given by\cirpol{
\begin{eqnarray}\label{DDscalartensor}
    \Omega _{st}(\eta ,k) &=& \frac{1}{108} \mathcal{A}_{\zeta}\mathcal{A}_{h}\frac{\Tilde{k}^{6}}{k^4}(16+24\Tilde{k}^2+\Tilde{k}^4) \overline{\mathcal{I}_{st}(k,v=\Tilde{k}^{-1},u=\Tilde{k}^{-1})^2}\Theta(2k_p-k) \text{.}
\end{eqnarray}}
\begin{figure} 
    \centering
    \includegraphics[width=0.7\textwidth]{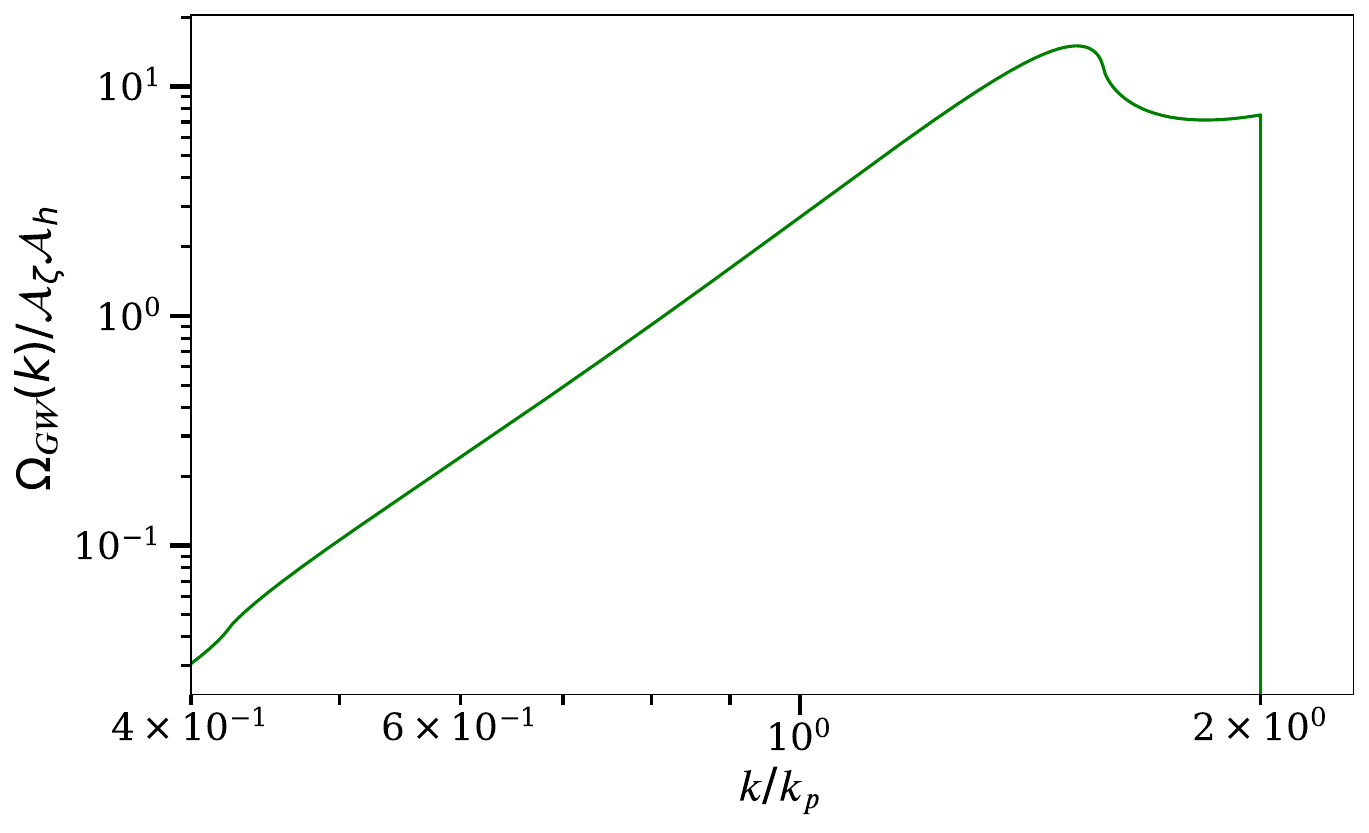}
    \caption{The spectral density of second order gravitational waves induced by scalar-tensor modes, where the primordial power spectra peak at the same scale $k_p$. The values needed to have a logarithmic resonance are allowed by momentum conservation but due to the structure of ($\ref{STfullkernel}$), the resonance does not occur. The amplitude spanned by the scalar-tensor induced waves is comparable to the scalar-scalar case.} 
    \label{DDst}
\end{figure}
Equation (\ref{DDscalartensor}) is shown in Fig.~\ref{DDst}. As expected, there is no resonant behavior.

\begin{figure} 
    \centering
    \includegraphics[width=0.7\textwidth]{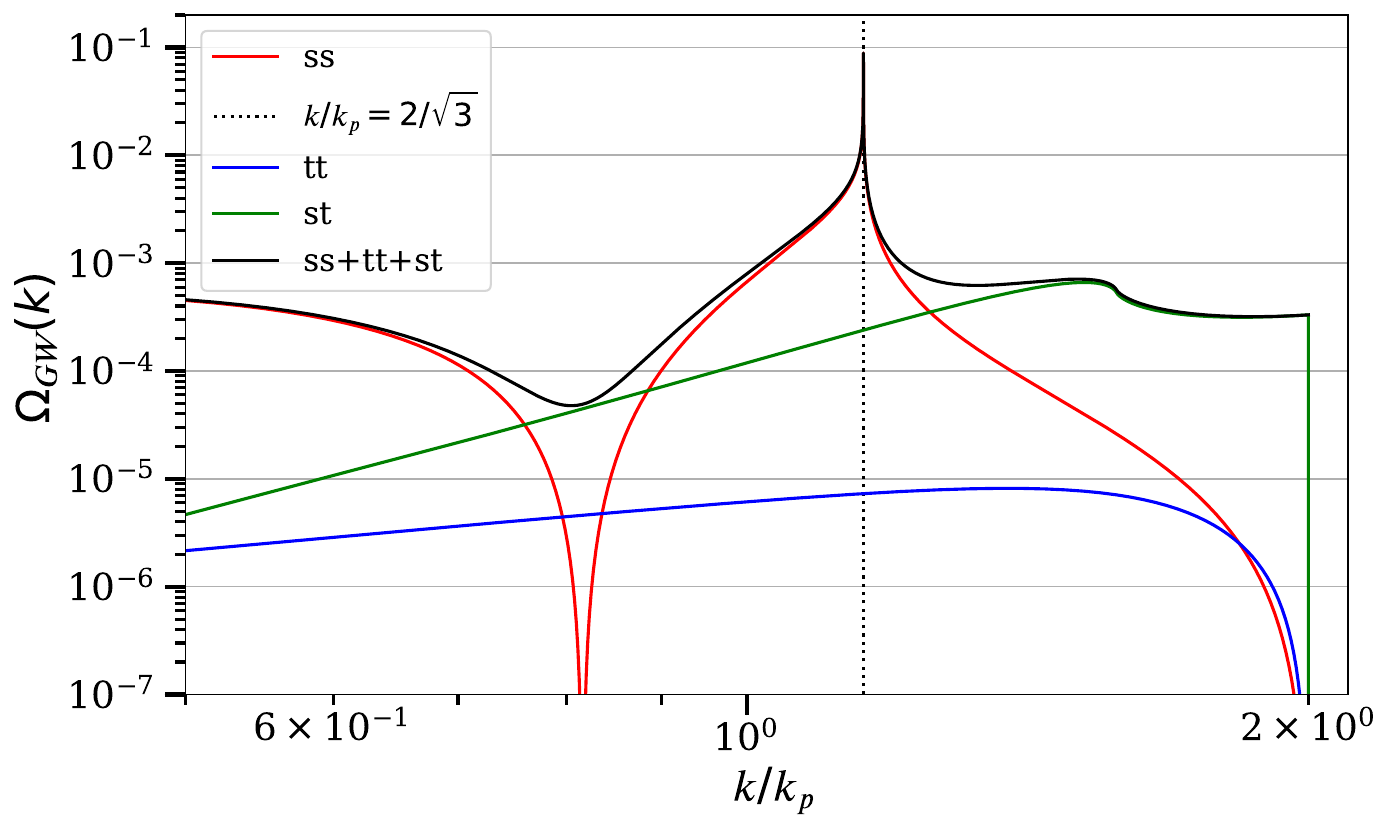}
    \caption{The total spectral density, where we have set $\mathcal{A}_{h}=0.1\mathcal{A}_{\zeta}$. The scalar-scalar (red line), tensor-tensor (blue line) and scalar-tensor (green line) contributions are shown. On small scales the tensor-tensor and scalar-tensor contributions become important, particularly the scalar-tensor mode of production dominates, as the amplitude of the scalar-scalar sourced waves decrease after the resonance. When considering the total contribution to the induced waves (black line) it becomes clear that for high values of $k$ the scalar-tensor contribution becomes the dominant mode of production for induced gravitational waves.}
    \label{DDssttst}
\end{figure}

Finally, we can compare all three contributions to the total spectral density, shown in Fig.~\ref{DDssttst}. In this plot, we see that the scalar-scalar contribution drops around six orders of magnitude between the resonant peak and the cut-off at $k=2k_{p}$, whilst the tensor-tensor and scalar-tensor contributions dominate, particularly the scalar-tensor one. This shows that tensor terms must be included in the analysis of induced gravitational waves when there is a peak in the primordial scalar and tensor power spectrum, as the scalar-tensor terms become the dominant mode of production.

We find that the total induced gravitational wave spectrum is in overall agreement with the results in Ref.~\cite{Chang:2022vlv} (see the blue curve in their Fig.~2). However, as far as the authors can tell we found that the amplitude of the spectral density drops off more steeply after the resonant peak on small scales.

\subsubsection{Peaks at different scales}
\begin{figure} 
    \centering
    \includegraphics[width=0.7\textwidth]{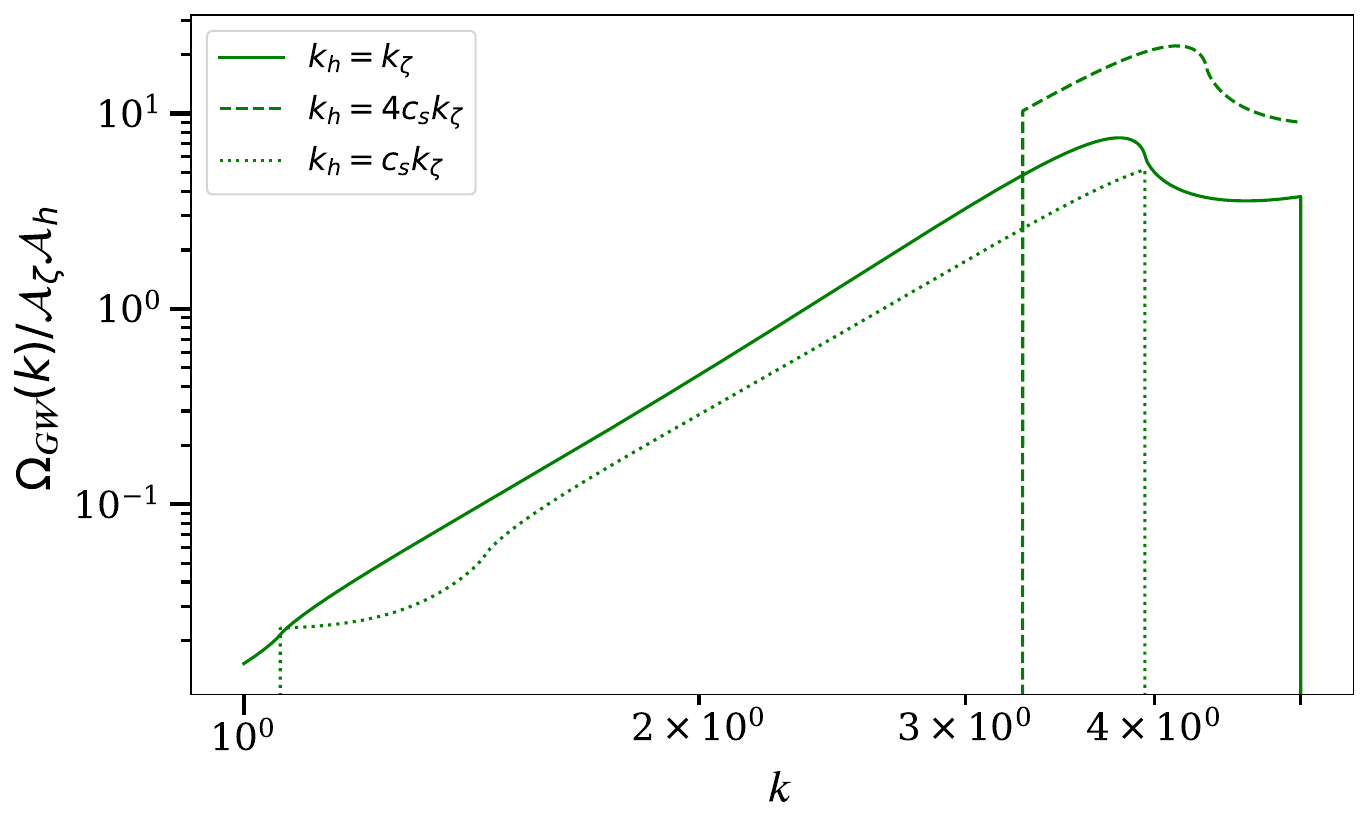}
    \caption{The spectral density of gravitational waves induced by scalar-tensor modes for a range of $k$ values. Here we present three different scenarios: when the primordial power spectra peak at the same scale (solid line), when the primordial tensor peaks on larger scales than the scalar-scalar resonant peak (dotted line), and when the primordial tensor peaks on smaller scales than the scalar-scalar resonant peak (dashed line). The `impulse' like behaviour of the dashed line can be understood from momentum conservation encoded in the Heaviside step functions.}
    \label{DD_st_dp}
\end{figure}

\begin{figure} 
    \centering
    \includegraphics[width=0.7\textwidth]{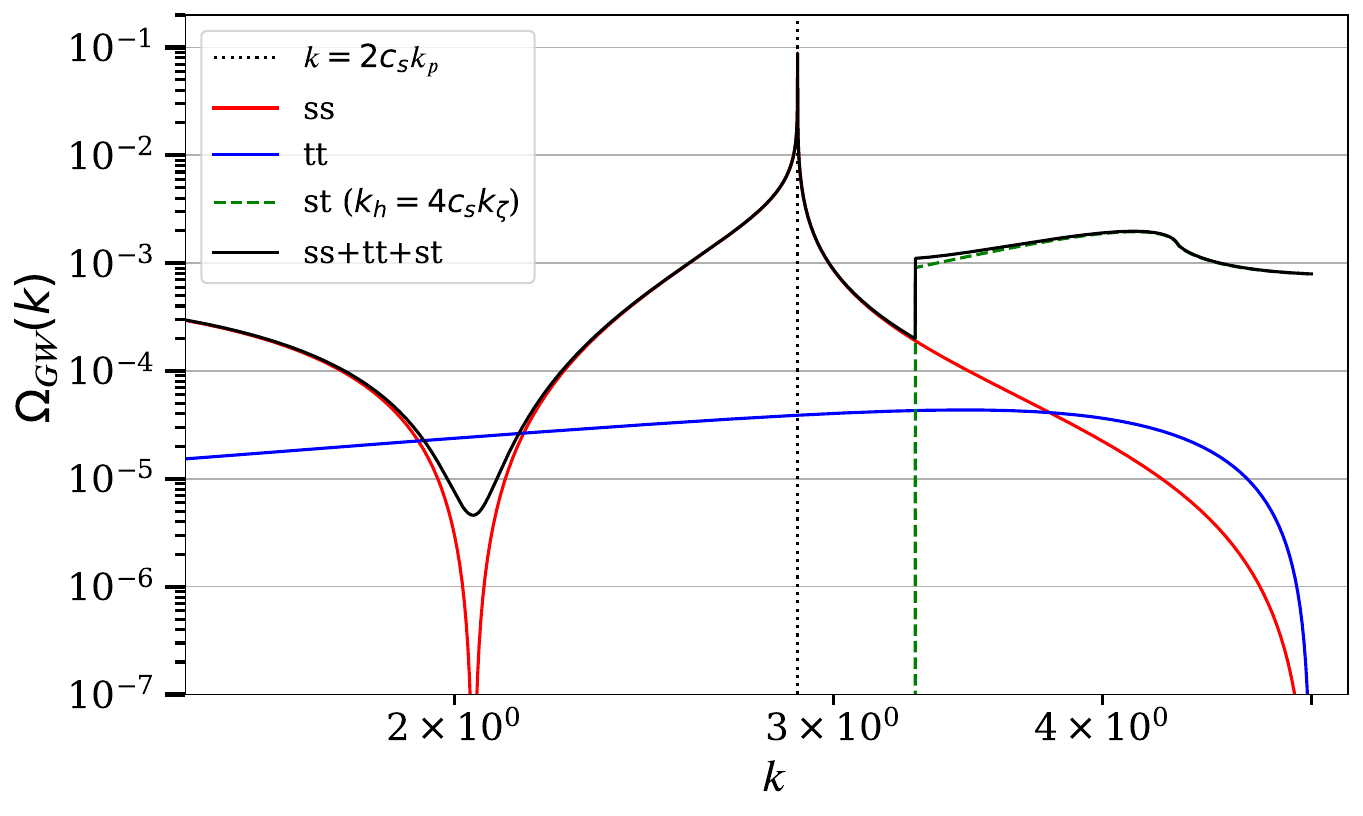}
    \caption{Full spectral density, where we have set $\mathcal{A}_{h}=0.1\mathcal{A}_{\zeta}$, with the scalar-tensor contribution (dashed green line) this time coming from a primordial power spectrum which peaks at a smaller scale than the scalar-scalar resonance.}
    \label{DD_ss_tt_st_dp}
\end{figure}

In this section we show how the second order scalar-tensor power spectrum changes when the primordial power spectra for scalars and tensors peak at different scales. The expression for the spectral density becomes
\cirpol{
\begin{eqnarray}\label{dd_dp_st}
    \Omega _{st}(k) &=& \frac{1}{216} \mathcal{A}_{\zeta}\mathcal{A}_{h}\frac{\left ((k-k_h)-k_{\zeta}^2 \right )^4+\left ((k+k_h)-k_{\zeta}^2 \right )^4}{k^6k_h^5k_{\zeta}} \overline{\mathcal{I}_{st}(k,v=k_h/k,u=k_{\zeta}/k)^2} \nonumber \\
    &&\times \Theta (k-|k_{\zeta}-k_h|) \Theta (-k+k_{\zeta}+k_h)\,,
\end{eqnarray}
}
where the two Heaviside step functions encode the integral limits in Eq.~(\ref{krangedirac}). We look at two different informative cases: first when the tensor power spectrum peaks at a scale higher than the resonance from the scalar-scalar contribution ($k_h\approx c_sk_{\zeta}$), and then at a scale smaller than the resonance ($k_h\approx 4c_sk_{\zeta}$). The results are presented in Fig.~\ref{DD_st_dp}. We see that when $k_h\approx c_sk_{\zeta}$ the scalar-tensor contribution is smaller than when the power spectra peak at the same scale, in fact we chose this value for $k_h$ because this value corresponds to an approximate threshold: when $k_h \leq c_sk_{\zeta}$ the signal will always be smaller than the case when $k_h=k_\zeta$. Secondly, we chose the peak scale $k_h\approx 4c_sk_{\zeta}$ to show that we can get higher amplitude signals than the previous setup. The short `burst' like signal is due to the allowed $k$ values encoded in the Heaviside step functions in Eq.~(\ref{dd_dp_st}) (\refrep{outside of this range, no incoming modes can satify momentum conservation and hence there is no outgoing mode}). Indeed, the closer $k_{h}$ is $k_{\zeta}$ the more the range of $k$ increases. In Fig.~\ref{DD_ss_tt_st_dp} we show all three contributions to the spectral density. The sharp feature in this signal is due to the fact that the input is a Dirac delta power spectrum.

\subsection{Lognormal Peak}
\begin{figure} 
    \centering
    \includegraphics[width=0.98\textwidth]{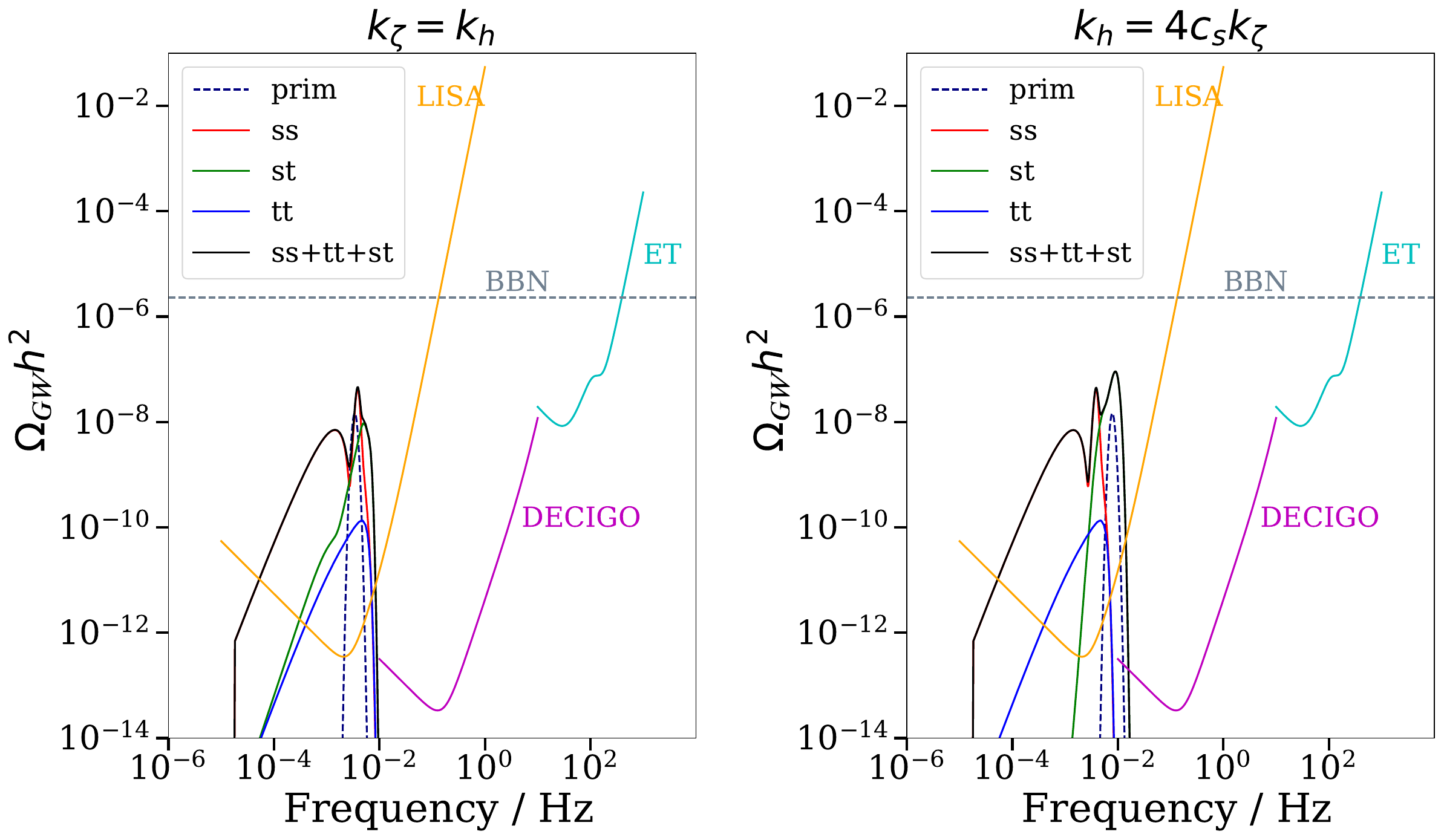}
    \caption{Plots of the spectral density against frequency. Each graph corresponds to a different setup in the location of the primordial tensor peak with $\mathcal{A}_{h}=0.1\mathcal{A}_{\zeta}$. We have also included the noise power spectral densities curves \cite{Moore:2014lga} for LISA, ET \cite{Sathyaprakash:2009xs} and DECIGO \cite{Kawamura:2020pcg}, as well as bounds on primordial gravitational waves from BBN \cite{LISACosmologyWorkingGroup:2022jok}. Furthermore, the dashed dark blue line is the primordial tensor power spectrum. We see that the scalar-tensor contribution is non-negligible. The sharp signals on larger scales and below the threshold for detection are artifacts of numerical integration.}
    \label{lognormaldetection_sp}
\end{figure}

As discussed earlier, a Dirac delta peak is not realistic, and hence to discuss the plausibility of detecting these induced waves we should move to a power spectrum which has some finite width to its peak. This can be achieved with a Lognormal power spectrum,
\begin{equation}
\label{lognorm_def}
    \mathcal{P}_{\zeta , h}= \frac{\mathcal{A}_{\zeta , h}}{\sqrt{2\pi}\sigma}\exp \left ( {-\frac{\log ^2(k/k_{\zeta , h})}{2\sigma ^2}}\right )\,,
\end{equation}
with $\sigma$ controlling the width of the peak and normalised such that $\int ^{\infty}_{- \infty}d\log k \mathcal{P}_{\zeta , h} = \mathcal{A}_{\zeta , h} $. In the following we take $\sigma =0.1$, so that the general behaviour of the spectral densities discussed in the previous section (Dirac delta input power spectra), holds in this scenario. For a detailed discussion of scalar-scalar induced waves from a lognormal peak, see Ref.~\cite{Pi:2020otn}. The scale at which we centre the scalar power spectrum ($k_{\zeta}$) is chosen so that the physical frequency is centred in the LISA band: $k=f \times (2\pi a_0)$, with $f_{\zeta}=3.4$ mHz. We take $\mathcal{A}_{\zeta}\approx 2.1\times10^{-2}$, which corresponds to an enhancement of $\mathcal{O}(10^{7})$ of the Planck 2018 data \cite{Planck:2018jri}, motivated by the enhancement needed to the power spectrum in order to produce primordial black holes \cite{Motohashi:2017kbs}. All spectral densities in the following are found by numerically integrating the expressions in Eqs.~(\ref{densityss}), (\ref{densitytt}) and (\ref{densityst}). In Fig.~\ref{lognormaldetection_sp} we present the results of our toy model for different scenarios. \refrep{When the primordial scalar and tensor spectra peak at the same scale, the scalar-tensor contribution becomes the dominant contribution. For the case when $k_h = 4c_sk_{\zeta}$ we can still distinguish the resonant peak and also have an extra contribution coming from the scalar-tensor terms, in the form of a peak.}

\section{Discussion}

In higher order cosmological perturbation theory lower order terms can couple to source the higher order terms. Hence second order  gravitational waves will be sourced by coupled first order scalar, vector, and  tensor terms. Until recently, research has  focused on source terms which are quadratic in first order scalar fluctuations, so-called SIGWs (see e.g. Ref.~\cite{Domenech:2021ztg}). In this paper, we investigated what happens if we include first order tensor fluctuations to contribute as a possible source. Two new sources appear: terms quadratic in tensor fluctuations and terms that couple a scalar and a tensor fluctuation. We have shown that the tensor-tensor contribution is subdominant due to its inherit structure whilst the scalar-tensor contribution is non negligible, particularly on small scales, in agreement with recent studies \cite{Chang:2022vlv, Bari:2023rcw, Yu:2023lmo}. Analytical results for the spectral density and kernels for the scalar-scalar, tensor-tensor and scalar-tensor contribution can be found in Sec. \ref{resultssecss}, \ref{resultssectt} and Sec. \ref{resultssecst} respectively.

In a $\Lambda$CDM universe, SOGWs sourced during the radiation dominated epoch of universe have a distinctive signal compared to the SIGWs. Due to the behaviour of source terms  including tensors in the UV limit, we studied the spectral density of waves induced by peaked spectra. In particular, we showed that  if there is peak in the first order tensor power spectrum, then the overall spectral density will be different to waves induced by quadratic scalar fluctuations, \refrep{particularly on small scales where the scalar-tensor waves can become dominant. Furthermore, we showed that when the primordial tensor power spectrum peaks at smaller scales than the scalar one, the spectral density gets an additional contribution which shows up as an additional peak (see Fig. \ref{lognormaldetection_sp}). Since the power spectrum for first order scalars and tensors is directly linked to inflation, our results can be used to constrain models of inflation which have a peak in the primordial tensor power spectrum.}

Work on second order  gravitational waves, sourced by first order scalar and tensor modes has so far focused mainly on toy model input power spectra. A natural next step is therefore to use more realistic input power spectra, both for the scalar and the tensor perturbations. In most cases this will no longer allow the use of analytical techniques and will require the use of numerical methods. Luckily there are already numerical packages to calculate the power spectrum for any single or multi-field model available, such as PyTransport \cite{Mulryne:2016mzv}. In a future extension of the work presented here we therefore plan to combine our results with PyTransport, which will allow us to further constrain the parameter space of models of inflation using gravitational waves.

\acknowledgments
The authors thank Chris Addis, Pedro Carrilho, Chris Clarkson, Charles Dalang, Matthew Davies, Pedro Fernandes, Laura Iacconi and David Mulryne for useful discussions. The authors are also grateful to Guillem Dom\`{e}nech for useful discussions and sharing an early draft of their work. RP is funded by STFC grant ST/P000592/1 and KAM is supported in part by STFC grants ST/T000341/1 and ST/X000931/1. \refrep{We would like to thank the anonymous referee for their feedback and useful comments.}

\paragraph{Note added.} Whilst in the final stages of this project, Ref.~\cite{Bari:2023rcw} and Ref.~\cite{Yu:2023lmo} were published on the arXiv.

\appendix
\newpage
\section{Evolution equations} \label{AppendixEvol}
Below we give the background, first order and second order relevant Einstein equations. We have used the $xPand$ package \cite{Pitrou:2013hga} for our computations. All calculations carried out below are done in the Newtonian gauge.
\subsection{Background equations}\label{AppendixEinsteinzeroth}
The energy-momentum tensor is given in Eq.~(\ref{EMtensor}). At $0^{th}$ order (background), the two Friedmann equations are
\begin{equation}
    \mathcal{H}^2 =\frac{a^2}{3M_{Pl}^2} \rho, \quad \mathcal{H}' = -\frac{a^2\rho}{6M_{Pl}^2}(1+3w), \quad \text{with} \quad P=w\rho \,,
\end{equation}
where  $w$ is the equation of state parameter. There is also a conservation equation
\begin{equation}
    \rho ^{\prime} + 3\mathcal{H}\left (\rho + P \right )=0\,.
\end{equation}
From the above we infer that during RD ($w=1/3$):
\begin{equation}
    a(\eta) \propto \eta, \quad \mathcal{H} = \frac{1}{\eta}\,.
\end{equation}
\subsection{First order equations} \label{1oEinstein}
The source terms in Eq.~(\ref{sourceterms}) have been simplified using first order relations we obtain from the Einstein equations, below we present only the relevant ones. The first order density perturbation and the scalar part of the velocity perturbation can be related to metric variables, from the $0-0$ and by taking the divergence of the $0-i$ first order Einstein equations respectively, we arrive to
\begin{eqnarray}
    \delta \rho &=& -2\rho \Phi + \frac{M_{Pl}^2}{a^2}\left (-6\mathcal{H}\Psi ^{\prime} +2\nabla ^2 \Psi \right )\,, \\
    v&=&-\frac{2}{3(1+w)\mathcal{H}} \left ( \Phi +\frac{\Psi ^{\prime}}{\mathcal{H}} \right )\,.
\end{eqnarray}
Furthermore, in the case of vanishing anisotropic stress, the symmetric trace-free part of the Einstein equations for a perfect fluid implies $\Phi = \Psi$. This simplifies the calculations significantly, namely we can get an evolution equation for $\Psi$ (the trace of the spatial part of the Einstein equations) in Fourier space:
\begin{equation}\label{firstorderscalareom}
     \Psi ^{\prime \prime} +3 \mathcal{H}\left (1+c_s^2  \right )\Psi ^{\prime} +  c_s^2 k^2\Psi  =0\,.
\end{equation}
The solution to this equation is shown in Eq.~(\ref{scalartransfer}). We have introduced the speed of sound:
\begin{equation}
    \delta P =c_s^2\delta\rho\,, 
\end{equation}
for adiabatic perturbations. We further note that when the equation of state is constant, then $c_s^2=w$, as is the case in our work.

Additionally, we are interested in the first order tensor evolution equation, using the TT projector, defined in Eq.~(\ref{TTprojector}), on the first order spatial Einstein equations we arrive at
\begin{equation}\label{firstordertensoreom}
    h_{ij}^{\prime \prime} + 2\mathcal{H}h_{ij}^{'} - \nabla ^2h_{ij} = 0\,.
\end{equation}
The solution to the first order tensor mode equation is shown in Eq.~(\ref{tensortransfer}).
\subsection{Second order equations}\label{2oEinstein}
The spatial part of the second order Einstein tensor is given by:
\begin{equation}
    \begin{split}
        G_{ij}^{(2)} &= h_{ij}^{\prime \prime (2)} + 2 \mathcal{H}h_{ij}^{\prime (2)} - h_{ij}^{(2)} (2\mathcal{H}^2 + 4\mathcal{H}^{\prime}) -\nabla ^2 h_{ij}^{(2)} + \delta _{ij} (\Phi ^{(2)} + \Psi ^{(2)})(2\mathcal{H}^2 + 4\mathcal{H}^{\prime}) \\
        &+2\delta _{ij} \Phi ^{\prime (2)}\mathcal{H} +4 \delta _{ij} \Psi ^{\prime (2)}\mathcal{H} + 2 \delta _{ij} \Psi ^{\prime \prime (2)} + \delta _{ij} \nabla ^2 \Phi ^{(2)} - \delta _{ij} \nabla ^2 \Psi ^{(2)}  \\
        &- \partial _i \partial _j \Phi ^{(2)} + \partial _i \partial _j \Psi ^{(2)} - 4 h_i^{m\prime}h_{jm}^{\prime} +  \delta _{ij} \left ( 3h_{mk}^{\prime} h^{mk \prime} + 4 h^{mk}h_{mk}^{\prime \prime} +8 \mathcal{H} h^{mk}h_{mk}^{\prime}  \right ) \\
        &- \Phi \left (4 h_{ij}^{\prime \prime} + 8 \mathcal{H}h_{ij}^{\prime} - 8\mathcal{H}^2 h_{ij} - 16\mathcal{H}^{\prime} h_{ij} \right ) - \delta _{ij} \left (8\mathcal{H}^2 + 16 \mathcal{H}^{\prime} \right ) \Phi ^{(2)} - 16 \delta _{ij} \mathcal{H} \Phi \Phi ^{\prime} \\
        &+\Phi ^{\prime} \left (8 \mathcal{H}h_{ij} - 2h_{ij}^{\prime} \right ) + \delta _{ij} \left ( 16 \mathcal{H}^{\prime} - 8\mathcal{H}^2 \right )\Phi \Psi - 8 \delta _{ij} \mathcal{H} \Psi \Phi ^{\prime} + 2 h_{ij}^{\prime} \Psi ^{\prime} + 24 \mathcal{H} h_{ij} \Psi ^{\prime} \\
        &-16 \delta _{ij} \mathcal{H} \Phi \Psi ^{\prime} - 4 \delta _{ij} \Phi ^{\prime} \Psi ^{\prime} + 2 \delta _{ij} \Psi ^{\prime 2} + 12h_{ij} \Psi ^{\prime \prime} - 8 \delta _{ij} \Phi \Psi ^{\prime \prime} - 4 \Psi \nabla ^2 h_{ij} + 4 h_i^m \partial _m \partial _j \Psi \\
        &+ \left ( 4 h_{ij} - 4\delta _{ij} \Phi \right )\nabla ^2 \Phi - \left ( 8 h_{ij} + 4\delta _{ij} \Psi \right ) \nabla ^2 \Psi + 4 h_j^m \partial _m \partial _i \Psi - 2 \partial _k h_{ij} \partial ^k \Phi \\
        & - 2 \delta _{ij} \partial _k \Phi \partial ^k \Phi - 6 \partial _k h_{ij} \partial ^k \Psi - 4 \delta _{ij} \partial _k \Psi \partial ^k \Psi + 4 h^{mk} \partial _m \partial _k h_{ij} - 4 h^{mk} \delta _{ij} \partial _m \partial _k \Phi \\
        & - 4 \partial _m h_{jk} \partial ^k h_i^m + 4 \partial _k h_{jm} \partial ^ k h_i^m - 4h^{mk}\delta _{ij} \nabla ^2 h_{mk} + 2 \delta _{ij} \partial _k h_{md} \partial ^d h^{mk} \\ 
        & - 3\delta _{ij} \partial _d h_{mk} \partial ^d h^{mk} + 2 \partial ^m \Phi \partial _i h_{jm} + 2 \partial ^m \Psi \partial _i h_{jm} - 4 h^{mk} \partial _i \partial _k h_{jm} + 2 \partial _i h^{mk} \partial _j h_{mk} \\
        &+2 \partial ^m \Psi \partial _j h_{im} +2 \partial ^m \Phi \partial _j h_{im} + 2\partial _i \Phi \partial _j \Phi - 2\partial _i \Psi \partial _j \Phi - 2\partial _i \Phi \partial _j \Psi +6 \partial _i \Psi \partial _j \Psi \\ 
        &-4h^{mk} \partial _j \partial _k h_{im} + 4 h^{mk} \partial _j \partial _i h_{mk} + 4\Phi \partial _j \partial _i \Phi + 4 \Psi \partial _j \partial _i \Psi
    \end{split}\,,
\end{equation}
where we have not used the background and first order equations for simplification yet. Similarly, the energy momentum tensor is given by:
\begin{eqnarray}
    T_{ij}^{(2)} &=& a^2 \bigg ( \delta _{ij} \left [ \frac{1}{2}P^{(2)} -2c_s^2\delta \rho \Psi -w\rho \Psi ^{(2)} \right]  + 2c_s^2\delta \rho  h_{ij} +(1+w)\rho \partial _i v \partial _j v \nonumber \\
    &+&  w\rho h_{ij}^{(2)}\bigg )\,.
\end{eqnarray}

\section{Fourier space conventions} \label{AppendixPol}
In this Appendix we present our choice of conventions in Fourier space in which we carry out our work.

A scalar quantity, $S(\mathbf{x},\eta)$, can be expressed as a Fourier integral
\begin{equation}
    S(\mathbf{x},\eta)=\frac{1}{2\pi^{\frac{3}{2}}} \int d^3\mathbf{k} S(\mathbf{k},\eta)e ^{i\mathbf{k}\cdot \mathbf{x}}\,.
\end{equation}
Similarly, a symmetric transverse-traceless tensor, $T_{ab}(\mathbf{x},\eta)$, in real space can be expressed as a Fourier integral
\begin{equation}
T_{ab}(\mathbf{x},\eta) =\frac{1}{2\pi^{\frac{3}{2}}} \int d^3\mathbf{k}  \left \{ T(\mathbf{k}, \eta) q_{ab}(\mathbf{k}) + \overline{T}(\mathbf{k}, \eta) \overline{q}_{ab}(\mathbf{k}) \right \}e ^{i\mathbf{k}\cdot \mathbf{x}} \,,
\end{equation}
where we note that the two polarizations decouple.
The tensor basis is given in terms of two time-independent polarization tensors $q_{ab}(\mathbf{k})$ and $\overline{q}_{ab}(\mathbf{k})$ which, respectively, are the `plus' and `cross' polarizations of a gravitational wave. These tensors can be expressed in terms of vectors that are orthogonal to the direction of propagation of the gravitational wave, $\mathbf{k}$:
\begin{equation}
\begin{split}
 q_{ab}^{+}(\mathbf{k})=\frac{1}{\sqrt{2}}\left ( e_a(\mathbf{k})e_b(\mathbf{k}) - \overline{e}_a(\mathbf{k}) \overline{e}_b(\mathbf{k})\right ) \\
 q_{ab}^{\times}(\mathbf{k})=\frac{1}{\sqrt{2}}\left ( e_a(\mathbf{k})\overline{e}_b(\mathbf{k}) + \overline{e}_a(\mathbf{k}) e_b(\mathbf{k})\right )
\end{split}
\end{equation}
The basis vectors obey:
\begin{eqnarray}
    e_a(\mathbf{k})e^a(\mathbf{k})=\overline{e}_a(\mathbf{k}) \overline{e}^a(\mathbf{k})=1, \quad e_a(\mathbf{k})\overline{e}^a(\mathbf{k})=0, \quad e_a(\mathbf{k})k^a=\overline{e}_a(\mathbf{k})k^a=0 \,,
\end{eqnarray}
and hence the symmetric tensor basis satisfies ($\lambda = +,\times$):
\begin{equation}
    q_{ab}^{\lambda}(\mathbf{k})\delta ^{ab} =0, \quad q_{ab}^{\lambda}(\mathbf{k}) k^a = 0, \quad q_{ab}^{\lambda}(\mathbf{k}) q^{ab,\lambda ^{\prime}}(\mathbf{k}) = \delta ^{\lambda \lambda ^{\prime}} \,.
\end{equation}
\cirpol{In this work we will be using circular polarizations, which are constructed from a linear combination of the plus and cross polarizations:
\begin{align}
    \begin{split}
        q^R_{ab}(\mathbf{k}) &= \frac{1}{\sqrt{2}} \left ( q^+_{ab}(\mathbf{k}) + i q^{\times}_{ab}(\mathbf{k})  \right ) \text{ ,}  \\
        q^L_{ab}(\mathbf{k}) &= \frac{1}{\sqrt{2}} \left ( q^+_{ab}(\mathbf{k}) - i q^{\times}_{ab}(\mathbf{k})  \right )\,,
    \end{split}
\end{align}
which are also transverse and traceless.  We choose the following normalisation condition
\begin{equation}
    \left( q_{ab}^{\lambda}(\mathbf{k}) \right)^* q^{ab,\lambda ^{\prime}}(\mathbf{k}) = \delta ^{\lambda \lambda ^{\prime}} \, ,
\end{equation}
for $\lambda=R,L$. Furthermore there are useful relations we use throughout our work
\begin{equation}
    \left ( q_{ab}^{\lambda}(\mathbf{k}) \right)^* = q_{ab}^{-\lambda}(\mathbf{k})  =q_{ab}^{\lambda}(-\mathbf{k})\,,
\end{equation}
where $-\lambda$ refers to the opposite polarization to $\lambda$, i.e. $-L=R$ and $-R=L$. }

It is useful to parameterize the vectors and basis tensors in spherical coordinates, since FLRW spacetime is spherically symmetric. The wave-vector $\mathbf{k}$ is expressed as: 
\begin{equation}
    \mathbf{k} = k \left (\sin{\theta _k} \cos{\phi _k}, \sin{\theta _k}\sin{\phi _k}, \cos{\theta _k} \right ) \,,
\end{equation}
in the $(k,\theta,\phi)$ coordinate system, with $\phi \in [0;2\pi[$ and $\theta \in [0;\pi]$. We can construct two orthonormal vectors in the subspace perpendicular to $\mathbf{k}$:
\begin{equation}
    \begin{split}
        &\mathbf{e}(\mathbf{k}) = \left ( \cos{\theta _k} \cos{\phi _k}, \cos{\theta _k}\sin{\phi _k}, -\sin{\theta _k} \right ) \,,\\
        &\overline{\mathbf{e}}(\mathbf{k}) = \left (-\sin{\phi _k}, \cos{\phi _k}, 0 \right )\,.
    \end{split}
\end{equation}
Furthermore, due to isotropy we have the freedom to align $\mathbf{k}$ with the $z$-axis, therefore $\theta _k = \phi _k =0$. Similarly, for the vector $\mathbf{p}$:
\begin{equation}
    \mathbf{p} = p \left (\sin{\theta _p} \cos{\phi _p}, \sin{\theta _p}\sin{\phi _p}, \cos{\theta _p} \right )\,,
\end{equation}
from which we get the important result:
\begin{equation}
    |\mathbf{k}-\mathbf{p}|^2 = k^2 + p^2 - 2kp\cos{\theta} _p\,.
\end{equation}

\subsection{\refrep{Setup for the tensor-tensor computation}}\label{setuptt}
In the definition of the polarization functions for the tensor-tensor contribution (Eq.~(\ref{polarazationdeftt1}) and Eq.~(\ref{polarazationdeftt2})) there is a `convoluted polarization tensor', $q_{ab}^{\lambda _1}(\mathbf{k}-\mathbf{p})$. For completeness, we clarify how we obtain expressions for this tensor in spherical coordinates.

In the source term quadratic in first order tensor perturbations, one of the wave-vectors of the tensor is fixed by the convolution. We call $\mathbf{q}$ the wave-vector of the convoluted first order tensor where $\mathbf{q}=\mathbf{k}-\mathbf{p}$ due to the convolution. In spherical coordinates, $\mathbf{q}$ reads 
\begin{equation}
    \mathbf{q} = q \left (\sin{\theta _q} \cos{\phi _q}, \sin{\theta _q}\sin{\phi _q}, \cos{\theta _q} \right ) \,,
\end{equation}
and we can solve for $q$, $\theta _q$ and $\phi _q$,
\begin{equation}
    q= \sqrt{k^2+p^2-2 k p \cos \theta _p }\text{,} \quad \theta _q = \arctan \left (\frac{p\sin \theta _p}{k-p\cos \theta _p} \right ) \text{ and} \quad \phi _q = \arctan \left ( \frac{\sin \phi _p}{\cos \phi _p} \right ) \text{.}
\end{equation}
Hence, the expression of the convoluted orthonormal vectors, $\mathbf{e}(\mathbf{k}-\mathbf{p})$ and $\overline{\mathbf{e}}(\mathbf{k}-\mathbf{p})$ are given by
\begin{equation}
    \begin{split}
        &\mathbf{e}(\mathbf{k}-\mathbf{p}) = \left (\frac{\left (-k+p\cos \theta _p \right )\cos \phi _p}{\sqrt{k^2 + p^2 - 2kp \cos \theta _p }} , \frac{\left (-k+p\cos \theta _p \right )\sin \phi _p}{\sqrt{k^2 + p^2 - 2kp \cos \theta _p }},- \frac{p\sin \theta _p}{\sqrt{k^2 + p^2 - 2kp \cos \theta _p }} \right ) \,,\\
        &\overline{\mathbf{e}}(\mathbf{k}-\mathbf{p}) = \left (\sin \phi _p,-\cos \phi _p, 0 \right )\,,
    \end{split}
\end{equation}
and they are used to construct $q_{ab}^{\lambda _1}(\mathbf{k}-\mathbf{p})$. 

\section{Polarization functions} \label{AppendixpolsGW2}
In this Appendix we show that the convolution integral is invariant under the switching of the two variables of the convolution and provide the results for the polarization function integrals in the azimuthal direction.

\subsection{Convolution symmetry in the tensor-tensor sector}\label{symmetryproof}
As stated previously, due to he nature of the second order computation (i.e. a convolution) there is a symmetry between $\mathbf{p}$ and $\mathbf{k}-\mathbf{p}$ that can be exploited to simplify computations when extracting the power spectrum for the scalar-scalar and tensor-tensor induced waves. This has been shown in Ref.~\cite{Malik:2006ir} and for the case of scalar-scalar induced waves in Ref.~\cite{Espinosa:2018eve}. Here we extend this reasoning for convoluted terms quadratic in tensor perturbations.

When we substitute Eq.~(\ref{ttcorrelation}) into Eq.~(\ref{twopointtensor}), we use the fact that integrating over $\delta (\mathbf{p}+\mathbf{p^{\prime}})$ or $\delta (\mathbf{p}+\mathbf{k^{\prime}}-\mathbf{p^{\prime}})$ is the same since there is a symmetry originating from the convolution. Consider a term that appears in Eq.~(\ref{GWexpression}), for example:
\begin{equation}\label{prooftensor}
    \int d^3\mathbf{p} q^{ab}_{\lambda}(\mathbf{k})q^{m,\lambda _1}_a(\mathbf{k}-\mathbf{p})q^{\lambda _2}_{bm}(\mathbf{p})f(p)f(k-p)\, .
\end{equation}
We want to show that the integral is unchanged by switching $\mathbf{p}$ and $\mathbf{k}-\mathbf{p}$. Let us consider the change of variable $\mathbf{\Tilde{p}}=\mathbf{k}-\mathbf{p}$, then the above changes to:
\begin{equation}
    \int d^3\mathbf{\Tilde{p}} q^{ab}_{\lambda}(\mathbf{k})q^{m,\lambda _1}_a(\mathbf{\Tilde{p}})q^{\lambda _2}_{bm}(\mathbf{k}-\mathbf{\Tilde{p}})f(k-\Tilde{p})f(\Tilde{p})\, .
\end{equation}
We have the freedom to relabel $a\leftrightarrow b$ and using the fact that $q^{ab}(\mathbf{k})$ is symmetric we have:
\begin{eqnarray}
    \int d^3\mathbf{\Tilde{p}} q^{ab}_{\lambda}(\mathbf{k})q^{m,\lambda _1}_b(\mathbf{\Tilde{p}})q^{\lambda _2}_{am}(\mathbf{k}-\mathbf{\Tilde{p}})f(k-\Tilde{p})f(\Tilde{p})\, .
\end{eqnarray}
This reasoning can be extended to all the tensor-tensor terms that appear in Eq.~(\ref{defpoldecomp}).
\subsection{Integral evaluations of the polarization functions} \label{azimuthalstuff}
\cirpol{
In this subsection we give analytical results of the polarization functions in the coordinates $v,u$. The following relations are used extensively
\begin{equation}
    \sin^2{\theta _p} = 1-\frac{(1+v^2-u^2)^2}{4v^2} = 1- \cos^2{\theta _p}\,.
\end{equation}
For the scalar-scalar part of the calculation we need to evaluate
\begin{equation}
    \int _0^{2\pi} d\phi _p \, q^{\lambda}_{ss}(\mathbf{k},\mathbf{p})q^{\lambda ^{\prime}}_{ss}(\mathbf{-k},\mathbf{-p}) = \frac{1}{2} k^4 \pi v^4 \left (1-\frac{(1+v^2-u^2)^2}{4v^2} \right )^2 \delta ^{\lambda \lambda ^{\prime}}  \, .
\end{equation}
The integration over the azimuthal plane ensures that the only non trivial contribution occurs when $\lambda = \lambda ^{\prime}$, which explains the factor of $2\pi\delta ^{\lambda \lambda ^{\prime}}$.
}

\cirpol{The tensor-tensor contribution is more involved, after integrating over $\mathbf{p^{\prime}}$ in Eq.~(\ref{ttcorrelation}) and expanding out Eq.~(\ref{twopointtensor}) we find that the quantity we need to evaluate is
\begin{align}
    \begin{split}
        &\sum_{\lambda _1, \lambda _2 , \lambda _1 ^{\prime}, \lambda _2 ^{\prime}} \int _0^{2\pi} d\phi _p \text{ } \bigg ( q^{\lambda , \lambda _1 , \lambda _2}_{tt,1}(\mathbf{k},\mathbf{p})q^{\lambda , \lambda _1 ^{\prime} , \lambda _2 ^{\prime}}_{tt,1}(\mathbf{-k},\mathbf{-p})\mathcal{I}^2_{tt,1}(x,v,u) +  q^{\lambda , \lambda _1 , \lambda _2}_{tt,2}(\mathbf{k},\mathbf{p})q^{\lambda , \lambda _1 ^{\prime} , \lambda _2 ^{\prime}}_{tt,2}(\mathbf{-k},\mathbf{-p}) \\ &\times \mathcal{I}^2_{tt,2}(x,v,u) +  q^{\lambda , \lambda _1 , \lambda _2}_{tt,1}(\mathbf{k},\mathbf{p})q^{\lambda , \lambda _1 ^{\prime} , \lambda _2 ^{\prime}}_{tt,2}(\mathbf{-k},\mathbf{-p})\mathcal{I}_{tt,1}(x,v,u)\mathcal{I}_{tt,2}(x,v,u) + q^{\lambda , \lambda _1 , \lambda _2}_{tt,2}(\mathbf{k},\mathbf{p}) \\
        &\times q^{\lambda , \lambda _1 ^{\prime} , \lambda _2 ^{\prime}}_{tt,1}(\mathbf{-k},\mathbf{-p})\mathcal{I}_{tt,1}(x,v,u)\mathcal{I}_{tt,2}(x,v,u)  \bigg ) \left (\delta ^{\lambda_1 \lambda_1^{\prime}}\delta ^{\lambda_2 \lambda_2^{\prime}} +\delta ^{\lambda_1 \lambda_2^{\prime}}\delta ^{\lambda_2 \lambda_1^{\prime}} \right ) \mathcal{P}_h^{\lambda _1}(p) \mathcal{P}_h^{\lambda _2}(|\mathbf{k}-\mathbf{p}|) \\
        &=2\times 2\pi \times \frac{1}{u^4v^4}  \bigg ( \frac{k^4}{32768} A(v,u)E^2(v,u)\mathcal{I}^2_{tt,1}(x,v,u) +\frac{k^4}{32768}D(v,u)H^2(v,u) \mathcal{I}^2_{tt,1}(x,v,u) \\
        & -\frac{k^2}{4096}A(v,u)E(v,u) \mathcal{I}_{tt,1}(x,v,u) \mathcal{I}_{tt,2}(x,v,u)-\frac{k^2}{4096}D(v,u)H(v,u) \mathcal{I}_{tt,1}(x,v,u) \mathcal{I}_{tt,2}(x,v,u)\\
        &+\frac{1}{2028}A(v,u)\mathcal{I}^2_{tt,2}(x,v,u) +\frac{1}{2028}D(v,u)\mathcal{I}^2_{tt,2}(x,v,u) + \frac{k^4}{65536} B(v,u)F^2(v,u)\mathcal{I}^2_{tt,1}(x,v,u) \\
        &+\frac{k^4}{65536} C(v,u)G^2(v,u)\mathcal{I}^2_{tt,1}(x,v,u) + \frac{k^4}{32768} I(v,u)G(v,u)F(v,u)\mathcal{I}^2_{tt,1}(x,v,u)  -\frac{k^2}{8192}\\
        &\times B(v,u)F(v,u) \mathcal{I}_{tt,1}(x,v,u) \mathcal{I}_{tt,2}(x,v,u) -\frac{k^2}{8192}C(v,u)G(v,u) \mathcal{I}_{tt,1}(x,v,u) \mathcal{I}_{tt,2}(x,v,u) \\
        & -\frac{k^2}{8192}I(v,u)F(v,u) \mathcal{I}_{tt,1}(x,v,u) \mathcal{I}_{tt,2}(x,v,u)  -\frac{k^2}{8192}I(v,u)G(v,u) \mathcal{I}_{tt,1}(x,v,u) \mathcal{I}_{tt,2}(x,v,u) \\
        &+\frac{1}{4056}B(v,u)\mathcal{I}^2_{tt,2}(x,v,u) +\frac{1}{4056}C(v,u)\mathcal{I}^2_{tt,2}(x,v,u)  +\frac{1}{2028}I(v,u)\mathcal{I}^2_{tt,2}(x,v,u)   \bigg ) \mathcal{P}_h(kv) \\
        &\times \mathcal{P}_h(ku) \delta ^{\lambda \lambda ^{\prime}}  \, .
    \end{split}
\end{align}
In the above, the first factor of two comes from taking $\mathcal{P}_h^R=\mathcal{P}_h^L$. We further note that there is an overall factor of $2\pi\delta ^{\lambda \lambda ^{\prime}}$, for reasons identical to the scalar-scalar case. Moreover, in order to reduce clutter, we have defined the functions $A(v,u)$, $B(v,u)$, $C(v,u)$, $D(v,u)$, $E(v,u)$, $F(v,u)$, $G(v,u)$, $H(v,u)$ and $I(v,u)$ which read} \cirpol{
\begin{subequations}
    \begin{align}
        A(v,u)&= (1+u-v)^2(1-u+v)^2(-1+u+v)^2(1+u+v)^6 \, ,
        \\
        B(v,u)&= (1+u-v)^2(1-u+v)^6(-1+u+v)^2(1+u+v)^2 \, ,
        \\
        C(v,u)&= (1+u-v)^6(1-u+v)^2(-1+u+v)^2(1+u+v)^2 \, ,
        \\
        D(v,u)&= (1+u-v)^2(1-u+v)^2(-1+u+v)^6(1+u+v)^2 \, ,
        \\
        E(v,u)&= 9u^2-2u(v-1)+(v+3)(v-1)\, ,
        \\
        F(v,u)&= 9u^2+2u(v-1)+(v+3)(v-1)\, ,
        \\
        G(v,u)&= 9u^2+2u(v+1)+(v-3)(v+1)\, ,
        \\
        H(v,u)&= 9u^2-2u(v+1)+(v-3)(v+1)\, ,
        \\
        I(v,u)&= \left ( (u-v)^2-1 \right )^4 (-1+u+v)^2(1+u+v)^2 \, .
\end{align}
\end{subequations}
}

\cirpol{
For the scalar-tensor calculation, we need to consider
\begin{align}
    \begin{split}
        &\sum_{\lambda _3, \lambda _3 ^{\prime}} \int _0^{2\pi} d\phi _p \text{ } q_{st}^{\lambda, \lambda _3}(\mathbf{k},\mathbf{p})q_{st}^{\lambda ^{\prime}, \lambda _3^{\prime}}(\mathbf{-k},\mathbf{-p}) \mathcal{P}_h^{\lambda _3}(p)\delta ^{\lambda_3 \lambda_3^{\prime}} \\
        &=2\times 2\pi \times \frac{1}{256v^4} \left ( (u^2-(1+v)^2)^4 + (u^2-(-1+v)^2)^4 \right )\mathcal{P}_h(kv) \delta ^{\lambda \lambda ^{\prime}} \, .
    \end{split}
\end{align}
}
\section{Evaluation of kernels}\label{Evaluationkernel}
\subsection{Definition of the functions}\label{Appendixsmallf}
In Eq.~(\ref{sourcetermsfourier}) we defined several functions. Here they are:
\begin{equation}
    \begin{split}
        f^{ss}(\eta ,k,|\mathbf{k}-\mathbf{p}|) &= \frac{2}{3(1+w)} \left [ \left ((T_{\Psi}(\eta |\mathbf{k}-\mathbf{p}|) + \frac{T^{\prime}_{\Psi}(\eta |\mathbf{k}-\mathbf{p}|)}{\mathcal{H}}\right )  \left ( T_{\Psi}(\eta p) + \frac{T^{\prime}_{\Psi}(\eta p)}{\mathcal{H}} \right )  \right ] \\
        &+T_{\Psi}(\eta |\mathbf{k}-\mathbf{p}|)T_{\Psi}(\eta p) \,,
    \end{split}
\end{equation}
tensor terms:
\begin{subequations}
    \begin{align}
        f^{tt}_1(\eta ,k,|\mathbf{k}-\mathbf{p}|) &=T_h(\eta |\mathbf{k}-\mathbf{p}|)T_h(\eta p)\,,
        \\
        f^{tt}_2(\eta ,k,|\mathbf{k}-\mathbf{p}|) &= T^{\prime}_h(\eta |\mathbf{k}-\mathbf{p}|)T^{\prime}_h(\eta p)\,,
    \end{align}
\end{subequations}
cross term:
\begin{eqnarray}
    f^{st}(\eta p,c_s \eta|\mathbf{k}-\mathbf{p}|) &=& -2p^2T_h(\eta p) T_{\Psi}(c_s \eta|\mathbf{k}-\mathbf{p}|)-\refrep{2(\mathbf{k}-\mathbf{p})\cdot \mathbf{p}}T_h(\eta p) T_{\Psi}(c_s \eta|\mathbf{k}-\mathbf{p}|)\nonumber \\
    &+& \mathcal{H}(1+3c_s^2)T_h(\eta p) T_{\Psi}^{\prime}(c_s \eta|\mathbf{k}-\mathbf{p}|)-|\mathbf{k}-\mathbf{p}|^2(1-c_s^2)T_h(\eta p)\nonumber \\
    &\times & T_{\Psi}(c_s \eta|\mathbf{k}-\mathbf{p}|)\,. 
\end{eqnarray}
We can write everything in terms of the variables $v$ and $u$ in Eq.~(\ref{uvcoordinates}). The scalar-scalar terms:
\begin{equation}
    \begin{split}
        f^{ss}(x,u,v,c_s) &=  \frac{2}{3(1+w)}  \left (T_{\Psi}(c_sxv) + x\frac{\partial T_{\Psi}(c_sxv)}{\partial x} \right )  \left ( T_{\Psi}(c_sxu) + x\frac{\partial T_{\Psi}(c_sxu)}{\partial x}  \right )   \\
        &+T_{\Psi}(c_sxv)T_{\Psi}(c_sxu)\,,
    \end{split}
\end{equation}
tensor-tensor terms:
\begin{subequations}
    \begin{align}
        f^{tt}_1(x,v,u) &=T_h(xv)T_h(xu)\,,
        \\
        f^{tt}_2(x,v,u) &= k^2\frac{\partial T_h(xv)}{\partial x}\frac{\partial T_h(xu)}{\partial x}\,,
    \end{align}
\end{subequations}
and finally the cross term:
\begin{eqnarray}
    f^{st}(x,u,v,c_s) &=& -2v^2k^2T_h(xv) T_{\Psi}(c_s xu)+\refrep{k^2(-1+u^2+v^2)}T_h(xv) T_{\Psi}(c_s xu)\nonumber \\
    &+&(1+3c_s^2)\frac{k^2}{x}T_h(xv) \frac{\partial T_{\Psi}(c_sux)}{\partial x} -u^2k^2(1-c_s^2)T_h(xv) T_{\Psi}(c_s xu) \,. \nonumber \\
\end{eqnarray}

\subsection{Kernel results}\label{kerneleval}

All the Kernels presented in Eq.~(\ref{ttkernel}) and Eq.~(\ref{stkernel}) can be evaluated analytically and our presented below. These are  trigonometric integrals, so it is useful to recall the following definitions:
\begin{equation}
    Si(z) = \int _0^z \frac{\sin t}{t} dt \, , \quad Ci(z) = -\int _z^{\infty} \frac{\cos t}{t} dt \, ,
\end{equation}
and their finite limits: $\lim_{z\to\infty} Si(z) = \pm\frac{\pi}{2}$ and $\lim_{z\to\infty} Ci(z) =0$.  

Firstly, we give the result from Ref.~\cite{Kohri:2018awv} for the scalar-scalar Kernel, $c_s=1/\sqrt{3}$:
\begin{eqnarray}
    I^{ss}(v,u,x\rightarrow\infty) &=& \frac{3(u^2+v^2-3)}{4u^3v^3x} \bigg [\sin{x} \left ( -4uv+(u^2+v^2-3) \log \bigg |\frac{3-(u+v)^2}{3-(u-v)^2} \bigg |\right ) \nonumber \\
    &&-\pi \cos{x}(u^2+v^2-3) \Theta (v+u-\sqrt{3})  \bigg]\,.
\end{eqnarray}

The evaluation of the tensor-tensor kernels are presented below. For concreteness, we show the steps for $\mathcal{I}_{c,1}^{tt}(x,v,u)$ and $\mathcal{I}_{s,1}^{tt}(x,v,u)$ in Eq.~(\ref{ttkernel}):
\begin{eqnarray}
\mathcal{I}_{c,1}^{tt}(x,v,u)&=&-\frac{1}{4uv} \bigg [\text{Si}((u-v+1) x)+\text{Si}((-u+v+1) x)+\text{Si}((u+v-1) x)\nonumber \\
&&-\text{Si}((u+v+1) x) \bigg ]\,,
\end{eqnarray}
\begin{eqnarray}
\mathcal{I}_{s,1}^{tt}(x,v,u)&=&\frac{1}{4uv} \bigg [\text{Ci}(|u-v-1| x)+\text{Ci}(|u-v+1| x)-\text{Ci}(|u+v+1|) x)\nonumber \\
&&-\text{Ci}(|u+v-1| x)-\log \bigg | \frac{(u-v-1)(u-v+1)}{(u+v+1)(u+v-1)} \bigg | \bigg ]\,.
\end{eqnarray}
Since we are interested in modes that are deep inside the horizon we take the limit where $k\eta >>1$, or equivalently sending $x\rightarrow \infty$.
\begin{eqnarray}
    \mathcal{I}_{c,1}^{tt}(x\rightarrow\infty ,v,u) = -\frac{\pi}{4uv}\Theta (u+v-1)\,,
\end{eqnarray}
\begin{eqnarray}\label{Itts1}
    \mathcal{I}_{s,1}^{tt}(x\rightarrow\infty ,v,u) = \frac{1}{4uv} \log \bigg | \frac{1-(u+v)^2}{1-(u-v)^2} \bigg |\,.
\end{eqnarray}
Similarly for $\mathcal{I}_{c,2}^{tt}(x,v,u)$ and $\mathcal{I}_{s,2}^{tt}(x,v,u)$: 
\begin{eqnarray}
\mathcal{I}_{c,2}^{tt}(x,v,u)&=& \frac{k^2}{2uvx^2} [ \sin{x}\sin{ux}\sin{vx}] + \frac{4k^2}{8uvx} [- v \sin{x} \sin{ux} \cos{xx}\nonumber \\
&&+ \cos{x} \sin{ux} \sin{vx}- u \sin{x} \cos{ux} \sin{vx} ] -
\frac{k^2 \left(u^2+v^2-1\right)}{8uv} \bigg [ \nonumber \\
&& \text{Si}((u-v+1) x)+\text{Si}((-u+v+1) x)-\text{Si}((u+v+1) x)\nonumber \\
&&-\text{Si}(-u x-v x+x)\bigg]\,,
\end{eqnarray}
\begin{eqnarray}
    \mathcal{I}_{c,2}^{tt}(x\rightarrow\infty ,v,u) &=& -\frac{\pi k^2 \left(u^2+v^2-1\right)}{8uv} \Theta(u+v-1)\,.
\end{eqnarray}
and
\begin{eqnarray}
\mathcal{I}_{s,2}^{tt}(x,v,u)&=&\frac{k^2}{2uvx^2} (\cos{x} \sin{ux} \sin{vx} ) + \frac{k^2}{2uvx} (\sin{x} \sin{ux} \sin{vx}+ v \cos{x} \sin{ux} \cos{vx} \nonumber \\
&&+ u \cos{x} \cos{ux} \sin{vx} ) - \frac{k^2}{8uv}\left(u^2+v^2-1\right) \bigg ( \text{Ci}(|u+v-1| x)+\text{Ci}(|u+v+1| x)\nonumber \\
&&-\text{Ci}(|u-v-1| x)-\text{Ci}(|u-v+1|) x +\log \bigg | \frac{(u-v-1)(u-v+1)}{(u+v-1)(u+v+1)} \bigg | \bigg  )\nonumber \\\,,
\end{eqnarray}
\begin{eqnarray}
    \mathcal{I}_{s,2}^{tt}(x\rightarrow\infty ,v,u) &=& \frac{k^2}{8uv} \bigg ( (u^2+v^2-1)\log \bigg | \frac{1-(u+v)^2}{1-(u-v)^2} \bigg | -4vu \bigg )\,.
\end{eqnarray}

Below we present the results for the scalar-tensor kernels, which is the evaluation of Eq.~(\ref{stkernel}). These expressions are much more involved than the tensor-tensor kernels, therefore we display only the part that will contribute to the final expression, i.e. the limit where $x\rightarrow \infty$.
$ \mathcal{I}_{c}^{st}(x\rightarrow\infty ,v,u)$ reads:
\begin{eqnarray}
    \refrep{\mathcal{I}_{c}^{st}(x\rightarrow\infty ,v,u)} &=& -\frac{\sqrt{3}\pi k^2}{16u^3v}\left ((u^2-3(v-1)^2)(u^2-3(v+1)^2) \right )\Theta (v+\frac{u}{\sqrt{3}}-1) \,, \nonumber \\
\end{eqnarray}
and $ \mathcal{I}_{s}^{st}(x\rightarrow\infty ,v,u)$:
\begin{eqnarray}
    \refrep{\mathcal{I}_{s}^{st}(x\rightarrow\infty ,v,u)} &=&\frac{ k^2}{16u^3v} \bigg (-\sqrt{3} (u^2-3(v-1)^2)(u^2-3(v+1)^2) \log \bigg | \frac{(\sqrt{3}v-u)^2-3}{(\sqrt{3}v+u)^2-3} \bigg |     \nonumber \\
    &+& 4uv\left (u^2-9v^2+9 \right ) \bigg ) \,.
\end{eqnarray}

\end{document}